%% file: tosem_arxiv.tex
\documentclass[acmsmall,screen,nonacm]{acmart}

\usepackage{amsmath,amssymb,amsfonts}
\usepackage{algorithmic}
\usepackage{graphicx}
\graphicspath{{./}}
\usepackage{textcomp}
\usepackage{xcolor}
\usepackage{booktabs}
\usepackage{url}
\usepackage{hyperref}
\usepackage{multirow}
\usepackage{xspace}
\usepackage{enumitem}
\usepackage{array}
\usepackage{longtable}
\usepackage{threeparttable}
\usepackage{tabularx}
\usepackage{microtype}
\setcopyright{none}
\renewcommand\footnotetextcopyrightpermission[1]{}
\acmJournal{TOSEM}
\acmYear{2026}
\acmVolume{0}
\acmNumber{0}
\acmArticle{0}
\acmMonth{8}
\newcommand{\codebleu}{\textsc{CodeBLEU}\xspace}
\newcommand{\passatk}{\textup{\texttt{pass@k}}\xspace}
\newcommand{\passat}[1]{\textup{\texttt{pass@#1}}\xspace}
\newcommand{\compileatk}{\textup{\texttt{compile@k}}\xspace}
\newcommand{\compileat}[1]{\textup{\texttt{compile@#1}}\xspace}

\newcommand{\fnz}{\texttt{fn0}\xspace}
\newcommand{\eclass}[1]{\textsc{#1}}
\newcolumntype{Y}{>{\raggedright\arraybackslash}X}
\input{artifact_release.tex}

\begin{document}

\title[Neural Decompilation of Dart AOT Binaries]{Evaluating Neural Decompilation of Dart AOT Binaries: Fine-Tuning, Metric Validity, Specification Leakage, and Reliability}
\renewcommand{\shortauthors}{Abualazm et al.}

\author{Raafat Abualazm}
\orcid{0009-0007-4843-5061}
\affiliation{%
  \institution{Cairo University}
  \department{Computer Engineering Department, Faculty of Engineering}
  \city{Giza}
  \country{Egypt}}
\email{raafat.202210476@eng-st.cu.edu.eg}

\author{Ayman AboElhassan}
\orcid{0000-0001-8703-4163}
\affiliation{%
  \institution{Cairo University}
  \department{Computer Engineering Department, Faculty of Engineering}
  \city{Giza}
  \country{Egypt}}
\email{ayman.abo.elmaaty@eng.cu.edu.eg}

\author{Amr G. Wassal}
\orcid{0000-0001-6009-4174}
\affiliation{%
  \institution{Cairo University}
  \department{Computer Engineering Department, Faculty of Engineering}
  \city{Giza}
  \country{Egypt}}
\email{wassal@eng.cu.edu.eg}

\renewcommand{\shortauthors}{Abualazm et al.}

\begin{abstract}
We present an execution-based evaluation of neural decompilation for Dart Ahead-of-Time (AOT) binaries and an audit of what its scores measure. Across six archived adapter--baseline comparisons, task-paired tests of \passatk at $k=1,5,10$ with Holm adjustment across 18 endpoints identify functional regressions in both Qwen3-8B adapters at every $k$. The other four comparisons are statistically inconclusive.

A separate common-candidate audit of three independently trained graph-prefix systems quantifies associations between static metrics and test-based correctness. On 141 reference-certified, contract-valid tasks, best \codebleu has modest task-level association with \passat{10} ($\rho=.218$--$.246$); \compileat{10} has weak association ($\rho=.072$--$.082$). Only 21.0--23.3\% of compiling candidates pass.

The input contract has a large within-run effect. In a paired single-seed intervention, removing semantic names and related cues as a bundle while retaining types, arity, and instruction content reduces coverage from 42/154 to 7/154 tasks. Separate matched graph-input perturbations produce no statistically detectable degradation under the semantic contract (six-test Holm $p\geq.750$); instruction-use attribution remains unresolved.

Across five decoding seeds on MF-174, the baseline solves 4.8 tasks on average, 15 at least once, and one in every seed, revealing solved-task turnover. Together, the studies provide a practical evaluation protocol: certify references, align metrics on common candidates, separate metadata and binary channels, repeat sampling, and preserve result provenance. The released capsule supports integrity checks and archived-outcome replay.
\end{abstract}

\ccsdesc[500]{Software and its engineering~Software reverse engineering}
\ccsdesc[300]{Computing methodologies~Neural networks}
\ccsdesc[300]{General and reference~Metrics}
\ccsdesc[300]{General and reference~Evaluation}

\keywords{neural decompilation, reverse engineering, empirical software engineering, large language models, Dart, benchmark validity, specification leakage, functional correctness, reproducibility}

\maketitle

\section{Introduction}
\label{sec:introduction}

\subsection{The Decompilation Challenge for Modern Languages}

Traditional decompilers such as Hex-Rays IDA Pro and Ghidra translate machine code into C-like pseudocode that is functionally informative but difficult to read. Compilation discards variable names, type annotations, comments, and high-level idioms~\citep{lacomis2019dire,cifuentes1994reverse}; the resulting generic identifiers and convoluted control flow impose a substantial burden during malware analysis, vulnerability discovery, program comprehension, and legacy-system maintenance.

The challenge is amplified for modern languages. Dart has sound null safety, Swift represents absence through optionals, and both feature closures, asynchronous control flow, object-oriented abstractions, and extensive runtime support. Dart's Ahead-of-Time (AOT) compiler additionally performs type specialization, inlining, tree shaking, and dead-code elimination. We study reconstruction of readable, functionally correct Dart from this production compilation mode. Swift provides a pragmatic AOT-compiled counterpart with comparable high-level abstractions and an available low-resource corpus for the training-mixture comparison. Its distinct compiler and \texttt{-O0} setting define a pipeline stress test. Alongside model performance, we examine whether the evaluation protocol measures the capability it is intended to assess.

\subsection{From Model Performance to Evaluation Construct Validity}

The study combines model comparison with evaluation validation. Six parameter-efficient adapters across three bases and two token-matched mixtures provide the historical comparison. Corrected functional tests identify regressions in the two Qwen3-8B adapters and leave the other four outcomes inconclusive. A separate same-candidate audit then measures how well static similarity and compilation track execution, extending the model-level comparisons on different panels in the submitted version.

An earlier public version described these patterns as significant Swift interference, capacity-dependent behavior, and a 200-instruction ``capability cliff.''\footnote{Earlier public version: \url{https://arxiv.org/abs/2607.06125}.} With the corrected functional inference and explicit treatment of the design constraints, we report the first two as training-mixture and model-configuration comparisons, and the length pattern as a descriptive association. Language and compiler optimization vary together, the bases differ in more than parameter count, and the instruction-count bins summarize observed tasks rather than locate a threshold.

A post-submission artifact audit brought the input contract into focus: \emph{what does the functional score measure when the benchmark supplies semantic metadata?} A descriptive function name, parameter names, helper symbols, a source path, or a demangled signature can act as a compact natural-language specification. Metadata-assisted reconstruction is useful in its own right. Distinguishing it from instruction-derived recovery requires controls over the information supplied with the binary.

We evaluate six complementary aspects of reconstruction:
\begin{enumerate}[leftmargin=*,itemsep=1pt,topsep=2pt]
    \item \codebleu measures static similarity to a reference implementation;
    \item \compileatk, under our harness, primarily measures interface conformance and type consistency;
    \item \passatk measures satisfaction of the available tests;
    \item matched metadata and binary-input controls test dependence on the supplied information channels;
    \item repeated sampling determines whether aggregate and task-level claims are stable;
    \item provenance controls determine whether the score corresponds to the intended prompt, data, checkpoint, and denominator.
\end{enumerate}

Symbol-bearing binaries define a legitimate \emph{metadata-assisted reconstruction} setting. Reporting that setting separately from metadata-scrubbed reconstruction makes the engineering task explicit and gives functional scores a clear interpretation.

A four-page preliminary report of the original fine-tuning study (RQ1--RQ4) appeared in the SANER 2026 ERA Track~\citep{abualazm2026llms}; it evaluated two bases and four adapters using \codebleu and compilation metrics only, with uncorrected dichotomized significance tests, and explicitly left execution-based semantic evaluation to future work. The historical \emph{generation data} predate that report and are not presented as newly collected experiments. The journal extension distinguishes reanalysis of retained evidence from new follow-up experiments:
\begin{enumerate}[leftmargin=*,itemsep=1pt,topsep=2pt]
\item \emph{Corrected scoring and statistics}: the third base, HE-154, and \passatk are added; the mislabeled any-pass McNemar analysis is replaced by paired task-level randomization tests at $k\in\{1,5,10\}$ with Holm correction jointly across all 18 variant--base endpoints (Section~\ref{sec:results});
\item \emph{Benchmark certification}: the reference-failure and contract-defect audits and the 150/142/141 denominators (Section~\ref{sec:benchmark_audit}) did not exist in the SANER report;
\item \emph{The construct-validity intervention} (semantic vs.\ typed-opaque vs.\ name-only) and the \emph{signature-only comparator} (Section~\ref{sec:construct_validity}) are entirely new experiments;
\item \emph{The common-candidate metric audit} on identical candidates from three independently trained seeds (Section~\ref{sec:aligned_metric_method}) is new;
\item \emph{The reliability study} (repeated-seed MF-174, Section~\ref{sec:sampling_reliability}), the \emph{input-channel interventions} (Section~\ref{sec:channel_integrated}), and the \emph{provenance post-mortem} (Section~\ref{sec:provenance}) are new studies on separately identified corpora;
\item \emph{Training-mixture and model-configuration comparisons} retain the measured contrasts and state their scope explicitly (Findings~2 and~3, Section~\ref{sec:results}).
\end{enumerate}
Table~\ref{tab:saner_delta} summarizes this same delta dimension by dimension.

\begin{table}[!t]
\centering
\caption{Feature-level extension beyond the four-page SANER ERA report.}
\label{tab:saner_delta}
\small
\begin{tabularx}{\columnwidth}{p{0.20\columnwidth}XX}
\toprule
Dimension & SANER ERA & This journal revision \\
\midrule
Models & 2 bases, 4 adapters & 3 bases, 6 adapters; external references separated \\
Evaluation & \codebleu and compilation & HE-154 \passatk; corrected paired inference at $k=1,5,10$ \\
Metric validity & Separate aggregate metrics & Three-seed, same-candidate association audit \\
Benchmark validity & Not studied & Reference certification, contract defects, metadata scrub, no-binary comparator \\
Reliability/lineage & Not studied & MF-174 reseeding, channel interventions, leakage post-mortem, claim ledger \\
Scope & x86-64 Dart examples & Additional preliminary FL-343 ARM64 infrastructure with explicit limits \\
\bottomrule
\end{tabularx}
\end{table}

\subsection{Research Questions}

The unified study addresses seven research questions.

\textbf{RQ1: Fine-tuning effectiveness.} Can small (4--8B) LLMs fine-tuned on assembly--Dart pairs improve functional correctness over their base models?

\textbf{RQ2: Data composition.} Under the two evaluated training pipelines, how do Dart-only and Dart--Swift mixtures compare?

\textbf{RQ3: Configuration dependence.} How do the observed mixture contrasts vary across the three tested base configurations?

\textbf{RQ4: Metric validity.} Do \codebleu and \compileatk track functional correctness measured by \passatk?

\textbf{RQ5: Benchmark construct validity.} How much of the observed functional score depends on semantic metadata rather than non-identifier binary evidence?

\textbf{RQ6: Sampling reliability.} How stable are aggregate \passatk counts and the identities of solved tasks across decoding seeds?

\textbf{RQ7: Information channels and provenance.} Which supplied channels affect interface conformance and correctness, and what evaluation controls are required for those effects to be interpretable?

RQ1--RQ3 concern the archived fine-tuning matrix. RQ4 combines its separate-panel comparisons with a direct common-candidate audit on a three-seed graph-prefix follow-up. RQ5--RQ7 concern post-submission analyses on separately identified corpora and checkpoints. Table~\ref{tab:study_matrix} identifies the design and role of each study; comparisons use the corresponding corpus and input contract.

\subsection{Contributions}

This article makes five contributions.

\begin{enumerate}[leftmargin=*,itemsep=2pt,topsep=2pt]
\item \textbf{An execution-based, multi-model Dart evaluation.} We evaluate six historical LoRA/DoRA variants across three base configurations and two token-matched data conditions using paired task-level inference. Corrected tests establish two configuration-specific functional regressions; the other four adapter comparisons are inconclusive. This separates measured changes in correctness from changes in static quality.
\item \textbf{An audited benchmark and metadata-sensitivity study.} We provide the historical 154-task benchmark and audited 150-, 142-, and 141-task subsets. A paired intervention establishes large within-run dependence on a semantic-metadata bundle. Distinct opaque name-only, unadapted signature-only/no-binary, and matched graph-perturbation conditions clarify which information each evaluation supplies.
\item \textbf{A direct common-candidate audit of metric validity.} A separate Qwen3-8B graph-prefix follow-up independently trains matched configurations at seeds 42--44 and records distinct checkpoints. Its aligned candidate pools quantify modest association between static similarity and correctness and weak discrimination by nearly saturated compilation coverage; execution identifies which compiling outputs satisfy the tests.
\item \textbf{A reliability and information-channel analysis.} Repeated-seed experiments quantify same-configuration variation and solved-task turnover. A separate channel study measures the effects of typed contracts, recovered constants, structural-payload substitution, and a no-binary oracle, distinguishing interface conformance from functional success.
\item \textbf{An auditable evaluation protocol and research record.} We connect benchmark certification, training/evaluation overlap checks, prompt-channel controls, sampling, and result provenance in a practical reporting protocol. The artifact records a historical scoring-test leak and its exclusions, and supports integrity checks and archived-outcome replay. Appendices retain exploratory architecture and objective-side investigations, a frontier-model reference point, and a preliminary ARM64 observation with their evidence status.
\end{enumerate}

\subsection{Paper Organization}

Section~\ref{sec:background} reviews Dart AOT decompilation, training, metrics, and related systems. Section~\ref{sec:overview} defines the corpora and evidence classes, and Section~\ref{sec:methodology} presents the training, evaluation, and audit procedures. Sections~\ref{sec:results} and~\ref{sec:error} report fine-tuning, metric, and complexity results. Sections~\ref{sec:construct_validity}, \ref{sec:sampling_reliability}, and~\ref{sec:channel_integrated} report construct validity, sampling reliability, and input-channel effects. Section~\ref{sec:provenance} documents result lineage and the leakage post-mortem. Sections~\ref{sec:discussion}--\ref{sec:conclusion} give the interpretation, limitations, recommendations, and conclusions. Appendices provide the full prompt audit and the secondary, exploratory, feasibility, and excluded investigations.

\section{Background and Related Work}
\label{sec:background}

\subsection{Dart AOT Compilation}

Dart supports Just-in-Time (JIT) compilation for development and Ahead-of-Time (AOT) compilation for deployment. AOT compilation applies type propagation, inlining, tree shaking, and dead-code elimination, producing optimized x86-64 native code. We study the production compilation mode, but the principal HE-154 benchmark still consists of short algorithmic functions rather than production application logic. We disable tree shaking using \texttt{@pragma('vm:entry-point')} annotations to preserve target boundaries for extraction. Existing Flutter reverse-engineering tools such as B(l)utter, reflutter, and JEB's Dart AOT support recover snapshots, symbols, and structural information but do not reconstruct original high-level source~\citep{blutter,reflutter,jebdartaot}.

\subsection{Parameter-Efficient Fine-Tuning}

Full fine-tuning of LLMs requires updating all parameters, which is computationally expensive. Low-Rank Adaptation (LoRA)~\citep{hu2022lora} addresses this by freezing the pretrained weights $W_0$ and injecting trainable rank decomposition matrices: $\Delta W = BA$, where $B \in \mathbb{R}^{d\times r}$, $A \in \mathbb{R}^{r\times k}$, and $r \ll \min(d,k)$. This reduces trainable parameters to under 1\% of the model while introducing no inference latency after merging ($W' = W_0 + BA$).

We employ Weight-Decomposed Low-Rank Adaptation (DoRA)~\citep{liu2024dora}, which decomposes the weight update into magnitude and direction components: $W' = m \cdot (W_0 + \Delta V)/\|W_0 + \Delta V\|_c$, where $\Delta V = BA$ is the LoRA update and $m$ is a trainable magnitude vector. DoRA more faithfully mimics full fine-tuning dynamics, particularly at low ranks~\citep{liu2024dora}.

\subsection{Evaluation Metrics}

\textsc{CodeBLEU}~\citep{ren2021codebleu} augments standard n-gram matching (BLEU) with two code-specific components: Abstract Syntax Tree (AST) matching and data-flow graph matching. We developed a custom \textsc{CodeBLEU} implementation for Dart.

\compileatk records whether at least one of $k$ candidates passes the Dart front end when concatenated with the real test harness. Under this harness, the tests invoke a named function with typed arguments and consume a typed return value. Compilation therefore requires name, arity, parameter and return-type conformance as well as body type checking. We consequently interpret \compileatk as an interface-conformance and type-consistency metric, not as a proxy for functional correctness.

\passatk measures the probability that at least one of $k$ generated outputs passes all unit tests, computed using the unbiased estimator: $\passatk = \mathbb{E}\left[1 - \binom{n-c}{k}/\binom{n}{k}\right]$, where $n$ is the total samples and $c$ is the number passing all tests~\citep{chen2021evaluating}. The growth of aggregate \passatk with $k$---the coverage curve---has been modeled as an inference scaling law under a simple statistical ansatz~\citep{levi2024inference}; the aggregate power law is consistent with per-task exponential decay under a heavy-tailed distribution of single-attempt success probabilities~\citep{schaeffer2025monkeys}; and beta-binomial models over per-task success probabilities can predict that curve from limited sampling budgets~\citep{kazdan2025efficient}.

\subsection{Classical Decompilation and Reconstruction}

Classical decompilation develops semantics-preserving structural analysis, control-flow recovery, and binary-analysis intermediate representations~\citep{schwartz2013native,yakdan2015no,brumley2011bap}. These systems aim to reconstruct program semantics from machine code, usually with synthesized identifiers. Neural approaches shift part of the problem toward source-code generation and readability, but they inherit the same requirement to distinguish recovered binary evidence from metadata supplied alongside it.

\subsection{Neural Decompilation}

\citet{katz2018using} established the sequence-to-sequence paradigm by applying RNNs to translate x86 assembly into C, and CODA~\citep{fu2019coda} developed an end-to-end neural program decompiler. SLaDe~\citep{armengol2024slade} trained a BART-style 200M-parameter sequence-to-sequence Transformer for assembly-to-C translation with robustness to compiler optimizations, while Nova~\citep{jiang2025nova} (ICLR 2025) built on DeepSeek-Coder with a hierarchical attention design specifically for assembly code structure, achieving 14.84--21.58\% absolute gains in Pass@1 and Pass@10 over prior work.

LLM4Decompile~\citep{tan2024llm4decompile} represented a scaling milestone, training DeepSeek-Coder models from 1.3B to 33B parameters on million-scale binary-source pairs and popularizing execution-based evaluation. The V2 series (LLM4Decompile-Ref, 2024) shifted to refinement: rather than end-to-end decompilation, models refine pseudo-code from Ghidra, with the 22B-V2 model outperforming the 6.7B-V1.5 by 40.1\% on re-executability. SK$^2$Decompile~\citep{tan2025sk2decompile} (October 2025) introduced a two-phase pipeline separating structural recovery (``skeleton'') from identifier naming (``skin'').

Recent systems add explicit structure or iterative refinement to the language model: control-flow augmentation~\citep{liu2025cfadecllm}, hierarchical graph abstraction~\citep{achamyeleh2026helios}, abstract logic trees~\citep{wang2025salt4decompile}, hierarchical assembly attention~\citep{jiang2025nova}, feedback-driven multi-turn refinement~\citep{liu2026autodecompiler}, and joint code/type prediction~\citep{dramko2025idioms}. These systems motivate the channel and architecture analyses preserved in the present article.

Complementary approaches apply LLMs to classical decompiler output. DeGPT~\citep{hu2024degpt} uses referee, advisor, and operator roles to refine Ghidra output; compilation checks alone do not establish preservation of semantics. Idioms~\citep{dramko2025idioms} jointly predicts code and type definitions. GENNM~\citep{xu2025unleashing} targets meaningful variable-name recovery from stripped binaries. Earlier refinement work~\citep{refining2023}, MDRE-LLM~\citep{mdrgllm2025}, and LLM-based deobfuscation~\citep{deobfuscation2026} address related reverse-engineering tasks. Studies of optimization levels and architectures in decompiled-code vulnerability analysis~\citep{vuln_analysis_decompiled2024} further motivate controlling these factors; our historical Dart--Swift experiment does not provide that control.

Broader surveys of LLMs for reverse engineering~\citep{sok_llm_re2025} and for source-code analysis~\citep{llm4sourcecode2025} corroborate this landscape. The systems discussed here predominantly target C, with examples in Go, Fortran, or OCaml~\citep{hosseini2022beyond}. Building on the Dart reconstruction study in our preliminary SANER report~\citep{abualazm2026llms}, this article adds execution-based functional evaluation and validity audits.

\subsection{Cross-Lingual and Multi-Language Code Models}

Source-to-source code translation includes unsupervised TransCoder for Java, C++, and Python~\citep{roziere2020transcoder} and multilingual program pretraining in PLBART~\citep{ahmad2021plbart}. Broader multilingual code generation and translation studies show that transfer varies with the language pair, task, training mixture, and model configuration~\citep{athiwaratkun2023multilingual,baltaji2025crosslingual,muennighoff2023crosslingual,cassano2023multiple}. Fixed-capacity multilingual pretraining can trade positive transfer against per-language capacity dilution~\citep{conneau2020unsupervised}; that literature motivates testing for negative transfer but does not explain a lower cross-language score by itself. Model-capacity studies likewise require controlled members of one family; parameter count alone is not identifiable when architecture, pretraining, distillation, and reasoning objectives also change~\citep{kaplan2020scaling}.

\citet{hosseini2022beyond} demonstrated retargetable decompilation to Go, Fortran, and OCaml by treating source and assembly as text, showing that multilingual decompilers are feasible without language-specific front-ends. Our low-resource mixture comparison uses Swift, another statically typed, AOT-compiled modern language with an available corpus reported as validated. The token-matched Dart-AOT/Swift-\texttt{-O0} design tests the resulting training pipelines; Section~\ref{sec:methodology} specifies their compiler and corpus differences.

\subsection{Evaluation Methodology in Decompilation}

Table~\ref{tab:positioning} shows the evaluation metrics used for performance evaluation in the related literature and positions our work against prior studies across key methodological dimensions.

Early work relied on token-level metrics (BLEU, exact match). SLaDe~\citep{armengol2024slade} adopted \textsc{CodeBLEU}~\citep{ren2021codebleu}, which augments n-gram matching with AST and data-flow components. Nova~\citep{jiang2025nova} reports Pass@1 and Pass@10. LLM4Decompile~\citep{tan2024llm4decompile} uses re-executability as a primary metric, requiring that decompiled code not only compile but also produce correct outputs.

DecompileBench (Gao et al., ACL Findings 2025)~\citep{gao2025decompilebench} is a comprehensive benchmark comprising 23,400 functions from 130 real-world programs evaluated across three dimensions: Recompile Success Rate (RSR), Coverage Equivalence Rate (CER), and LLM-judged code quality. Their finding that LLM-based decompilers surpass commercial tools in understandability despite 52.2\% lower functional correctness illustrates why surface quality and functional correctness need separate measurement. Our aligned audit measures their relationship directly in the Dart setting.

Execution-based decompilation evaluation appears in several forms. LLM4Decompile's re-executability compares outputs of generated and original programs on selected inputs, while HumanEval-style \passatk evaluates multiple stochastic candidates against a task harness. These protocols differ in test construction and coverage; neither alone establishes comprehensive semantic equivalence. \citet{dramko2024taxonomy} document fidelity problems in surface-only C-decompiler evaluation, reinforcing the need to report the exact executable contract. We extend this execution-based tradition with HumanEval-Dart harnesses for 154 historical Dart tasks, benchmark certification, and a direct metric audit. Section~\ref{sec:benchmark_audit} distinguishes contract-valid, reference-certified, and strict-clean subsets.

The Self-Constructed Context Decompilation work of \citet{feng2024self} combines fine-grained source-to-assembly alignment (using debug information) with a self-improving loop that recompiles model outputs and re-injects them as demonstrations. Its structured supervision is relevant to the instruction-count association reported here, while its execution feedback is a distinct future training direction.

\begin{table}[!t]
\centering
\caption{Positioning against prior neural decompilation studies. CB = \textsc{CodeBLEU}; c@k = \texttt{compile@k}; p@k = \texttt{pass@k}; Stats = paired statistical tests. X-ling marks inclusion of cross-language training or mixture comparisons, not identification of a causal language effect. Nova reports Pass@1/Pass@10. Explicit labels denote related metrics rather than the column's exact $k$-sample endpoint: R$^\ast$ = re-executability; RSR = Recompile Success Rate; CER = Coverage Equivalence Rate.}
\label{tab:positioning}
\setlength{\tabcolsep}{3.5pt}
\small
\begin{tabular}{lllccccc}
\toprule
Study & Lang. & Size & CB & c@k & p@k & X-ling & Stats \\
\midrule
Katz~\citep{katz2018using} & C & RNN & & & & & \\
SLaDe~\citep{armengol2024slade} & C & 200M & \checkmark & & & & \\
Nova~\citep{jiang2025nova} & C & 1B & & & \checkmark & & \\
LLM4Dec.~\citep{tan2024llm4decompile} & C & 33B & \checkmark & \checkmark & R$^\ast$ & & \\
Hosseini~\citep{hosseini2022beyond} & Multi & Var. & & & & \checkmark & \\
DeGPT~\citep{hu2024degpt} & C & LLM & \checkmark & & & & \\
Idioms~\citep{dramko2025idioms} & C & --- & & \checkmark & & & \\
GENNM~\citep{xu2025unleashing} & C & 2--34B & & & & & \\
DecBench~\citep{gao2025decompilebench} & C & Var. & & RSR & CER & & \\
\midrule
\textbf{This work} & \textbf{Dart} & \textbf{4--8B} & \checkmark & \checkmark & \checkmark & \checkmark & \checkmark \\
\bottomrule
\end{tabular}
\end{table}

\subsection{Semantic Metadata and Prompt-Channel Controls}

Neural-decompilation prompts can expose semantic metadata through several channels: an explicit source signature, parameter names, demangled symbols embedded in disassembly, decompiler-inferred declarations, or source-path fields. We audited six published evaluation configurations across four systems for which prompt templates, construction code, or released evaluation data were available. The detailed protocol and row-level evidence appear in Appendix~\ref{app:prompt_audit}; Table~\ref{tab:prompt_audit_summary} summarizes the result.

\begin{table*}[!t]
\centering
\caption{Semantic-metadata channels in six auditable neural-decompilation evaluation configurations. ``Control'' means a metadata-scrubbed arm or signature-only baseline.}
\label{tab:prompt_audit_summary}
\scriptsize
\setlength{\tabcolsep}{2.5pt}
\begin{tabularx}{\textwidth}{Y l l Y c c}
\toprule
System / split & Name & Parameters & Signature channel & Control & Seeds \\
\midrule
L4D-End~\citep{tan2024llm4decompile} / HumanEval-Dec. & \texttt{func0} & --- & --- & no & 1 \\
L4D-Ref~\citep{tan2024llm4decompile} / HumanEval-Dec. & \texttt{func0} & \texttt{param\_N} & Ghidra & no & 1 \\
L4D~\citep{tan2024llm4decompile} / GitHub2025, ExeBench & original & --- & C++ types & no & 1 \\
SALT4Dec.~\citep{wang2025salt4decompile} / Dec.-Eval, MBPP & \texttt{func} & \texttt{var1..} & --- & no & 1 \\
SK2Dec.~\citep{tan2025sk2decompile} / stripped & \texttt{sub\_XXXX} & \texttt{a1,a2} & IDA & no & 1 \\
SCDBench~\citep{qin2026scdbench} / EVM-Solidity & original & --- & types + arity & no & 1 \\
\bottomrule
\end{tabularx}
\end{table*}

None of the six audited configurations reports a metadata-scrubbed arm or signature-only baseline. Exposure also varies by benchmark split: HumanEval-Decompile uses the semantically empty symbol \texttt{func0}, whereas other splits evaluated through the same prompt path retain true symbols or demangled signatures. This split-level variation motivates separate metadata-assisted and instruction-derived evaluation conditions. The audit describes the six verifiable configurations, not prevalence across all modern-language evaluations.

\subsection{Positioning of This Work}

The study brings seven evaluation elements together for Dart AOT decompilation: (1) a modern-language AOT target; (2) execution-based functional evaluation; (3) token-matched same-language and cross-lingual fine-tuning; (4) paired analysis across several base capabilities; (5) separate measures of static similarity, interface conformance, and functional correctness; (6) metadata-scrubbed conditions and a signature-only comparator; and (7) repeated-seed and provenance-aware evaluation.

DecompileBench~\citep{gao2025decompilebench} independently shows that understandability and runtime consistency can diverge on real C programs. Our historical fine-tuning matrix shows differing aggregate movements on separate panels, and the common-candidate audit quantifies metric association directly. The metadata intervention adds a further result: the supplied specification can substantially affect execution success. Together, these studies connect model performance to the information and tests that define the reconstruction task.

\section{Study Overview, Corpora, and Evidence Taxonomy}
\label{sec:overview}

The research programme combines historical model comparisons, direct metric and metadata audits, and sampling and channel studies. Table~\ref{tab:study_matrix} identifies each study's corpus, model family, seed design, and evidentiary role. Numerical comparisons are made within studies, since the rows use different evaluation conditions.

\begin{table*}[!t]
\centering
\caption{Study matrix. The studies use different corpora, models, prompts, toolchains, and seed designs. Agreement across rows is conceptual convergence, not direct replication.}
\label{tab:study_matrix}
\scriptsize
\setlength{\tabcolsep}{2pt}
\begin{tabularx}{\textwidth}{>{\raggedright\arraybackslash}p{0.115\textwidth}>{\raggedright\arraybackslash}p{0.095\textwidth}>{\raggedright\arraybackslash}p{0.18\textwidth}>{\raggedright\arraybackslash}p{0.105\textwidth}>{\raggedright\arraybackslash}p{0.14\textwidth}Y}
\toprule
Study & Corpus & System & Seeds / samples & Evidence class & Principal role \\
\midrule
Fine-tuning & CB-73, CP-126, HE-154 & Qwen3-4B, DeepSeek-R1-Qwen3-8B, Qwen3-8B and six adapters & one training run and one principal generation set per configuration; $n=10$ on HE & \eclass{Primary, configuration-specific} & RQ1--RQ3: observed adaptation and mixture contrasts \\
Training-seed replication and aligned metric audit & HE-154 and HE-141 & Qwen3-8B graph-prefix configurations & independent training seeds 42--44, distinct checkpoints; the same 10 candidates/task for every metric & \eclass{Replicated secondary, provenance-limited} & RQ4 and robustness: seed variation plus task- and candidate-level metric association \\
Semantic-metadata/name-only intervention & HE-154 and audited subsets & Qwen3-8B + graph encoder, frozen checkpoint & seed 42; 10 candidates/task & \eclass{Primary, single-seed} & RQ5: semantic, typed-opaque, and opaque-name-only contracts \\
Signature-only comparator & HE-154 & pretrained, unadapted base and three binary-conditioned systems & one paired archived evaluation & \eclass{Secondary} & No-binary performance reference; adaptation also differs \\
Graph-input ablation & Semantic and opaque name-only views of HE-154 & frozen Qwen3-8B graph-prefix checkpoint & seed 42; 10 candidates/task in each of six runs & \eclass{Secondary, single-seed} & RQ5: verified instruction shuffle and cross-task substitution; opaque floor limits inference \\
Reseeding and task-set stability & MF-174 & T5Gemma-2-4B-4B, frozen checkpoint & seeds 42--46 & \eclass{Replicated secondary} & RQ6: same-configuration variation and task identity turnover \\
Input-channel interventions & MF-174 & same frozen checkpoint & five blocks; selected arms extended to seed 47 & \eclass{Secondary / exploratory} & RQ7: typed contract, constants, structure, and no-binary oracle \\
Cross-system, objective, architecture, and frontier diagnostics & HE-154, MF-175 & multiple systems & mixed & \eclass{Exploratory / feasibility} & Preserved in appendices; no main causal claim \\
ARM64 Flutter observation & FL-343 & encoder--decoder, generated Dart in release APKs & one seed & \eclass{Preliminary} & Cross-ISA and length-scaling observation only \\
Historical leaked pools & HE-154 & historical prompt builder & mixed & \eclass{Withdrawn} & Provenance record; excluded from findings \\
\bottomrule
\end{tabularx}
\end{table*}

\subsection{Corpora}

\begin{table}[!t]
\centering
\caption{Evaluation corpora and the claims they support.}
\label{tab:corpora_integrated}
\small
\begin{tabular}{llrl}
\toprule
ID & Origin & Tasks & Role \\
\midrule
CB-73 & held-out Dart & 73 & \codebleu study \\
CP-126 & held-out Dart & 126 & \compileatk study \\
HE-154 & HumanEval ported to Dart & 154 & historical \passatk and metadata audit \\
MF-175 & internal generated + legacy Dart & 175 & reseeding and channel studies \\
FL-343 & obfuscated Flutter ARM64 APKs & 343 & preliminary cross-ISA observation \\
\bottomrule
\end{tabular}
\end{table}

\textbf{HumanEval-Dart.} HE-154 comprises 154 of HumanEval's 164 problems with Dart harnesses (median seven direct candidate assertions; range 1--26), in the benchmark-porting tradition of MultiPL-E~\citep{cassano2023multiple}. A retained retrospective author account describes asking ChatGPT-5.5 with \texttt{xhigh} reasoning to translate and execute the tasks for correctness checks. The original translation prompt, run transcript, and task-specific rejection records are unavailable; Section~\ref{sec:benchmark_audit} distinguishes this reported procedure from the current certification. The retained ID set equals MultiPL-E's 157-task Dart set minus tasks 23, 25, and 162, with seven other HumanEval tasks absent from both sets. Appendix~\ref{app:benchmark_audit} documents this membership correspondence without inferring local selection decisions or code lineage. Independent execution of the retained references finds twelve failures, leaving 142 reference-certified tasks; four tasks have contract or test defects, leaving 150 contract-valid tasks; their intersection is the strict-clean set of 141. Historical tables retain the legacy 154-task denominator, while the metadata intervention reports all four denominators.

\textbf{MF-175.} MF-175 is an internally constructed, held-out x86-64 suite, not an external benchmark. It contains 136 LLM-generated tasks and 39 legacy programs with generated differential tests. One evaluation task shares byte-identical acceptance tests with a training row and is excluded from the replicated analyses, producing MF-174. The retained legacy records do not identify original repositories or licences, and the majority-generated composition limits external validity. The public artifact therefore distinguishes redistributable derived results from raw materials whose provenance remains unresolved.

\textbf{FL-343.} FL-343 consists of 343 functions compiled into obfuscated, signed Flutter release APKs targeting ARM64. It extends the evaluation infrastructure to genuine release-mode ARM64 artifacts. The sources are model-generated, and the existing run retains semantic names and uses one seed, making it a preliminary cross-ISA observation rather than a replication on non-synthetic application code.

\subsection{Evidence Classes}

We use five labels throughout the article and appendices.

\begin{description}[leftmargin=2.7cm,style=nextline]
\item[\eclass{Primary}] Directly supports a principal conclusion under the stated design.
\item[\eclass{Replicated secondary}] Multi-seed evidence on a separate corpus or system that refines, but does not replicate, a primary result.
\item[\eclass{Exploratory}] Informative evidence limited by power, matching, multiplicity, corpus composition, or retrospective selection.
\item[\eclass{Feasibility}] An attempted method could not produce the intended measurement; examples include out-of-memory RL launches and an unavailable distillation interface.
\item[\eclass{Withdrawn}] Retained for lineage and audit, but excluded from scientific claims because of leakage or failed provenance.
\end{description}

The labels distinguish the evidence used for the main findings from the exploratory, feasibility, and excluded records retained for audit.

\section{Methodology}
\label{sec:methodology}

\subsection{Training Data}

\subsubsection{Natural data.}
The foundation of the fine-tuning corpus consists of 246 natural Dart functions sourced from RosettaCode projects, representing diverse algorithmic patterns.

\subsubsection{Synthetic data generation.}
Multiple LLM families (GPT-4, Claude, DeepSeek, and Qwen) generated 948 synthetic Dart examples and 754 synthetic Swift examples to augment the natural data. The submitted run report describes compilation filtering, deduplication, and filtering at 100 tokens. The retained aggregate \texttt{dart\_all.jsonl} has 1,195 rows rather than the reported 1,194, including 1,007 distinct source strings and 188 duplicate-source rows; it also contains assembly-duplicate groups. We distinguish that aggregate snapshot from the reported training corpus because the exact derived-corpus manifest for the six runs was not archived.

The aggregate Dart file contains non-empty reasoning text in 1,193/1,195 rows (median 359 words). Records omit the generating model revision, prompt, seed, filtering decision, and validation provenance. Public candidate trainers implement different supervision paths: one uses code only, while others serialize \texttt{<think>} reasoning followed by fenced source. Inference templates can request reasoning; scoring uses only extracted Dart code. No run-to-script ledger identifies the path used for each of v1--v6, so the exact chain-of-thought training and inference configuration remains unresolved and is not used to explain the observed effects.

\subsubsection{Choice of Swift and token-matched conditions.}
Swift provides a low-resource mixture stress test: it shares static typing, AOT compilation, closures, generics, and substantial runtime abstractions with Dart while using a distinct compiler and toolchain.

The two training conditions were:
\begin{itemize}
    \item \textbf{Dart + synthetic Dart:} 1,194 assembly--Dart pairs (246 natural + 948 synthetic);
    \item \textbf{Dart + Swift:} 1,000 assembly--source pairs (246 Dart + 754 Swift).
\end{itemize}
The total token counts are matched to within 2\%. This controls data volume but not compiler optimization: Dart uses production AOT optimization, whereas Swift is compiled at \texttt{-O0}. Section~\ref{sec:threats} treats that mismatch as a major confound for RQ2--RQ3.

\subsection{Assembly Generation}
\label{sec:assembly_generation}

Dart source was compiled using \texttt{dart compile aot-snapshot}, which applies type propagation, inlining, dead-code elimination, and related AOT optimizations. Tree shaking was disabled through \texttt{@pragma('vm:entry-point')} to preserve target function boundaries. Swift source was compiled at \texttt{-O0} to preserve program structure.

We report four distinct size measures: \emph{source lines} in the Dart reference, \emph{raw disassembly lines} (non-empty tool-output lines, including headers and symbol declarations), \emph{parsed instruction count} (address--offset machine-instruction records), and \emph{tokenizer tokens}. Unless stated otherwise, length analyses use parsed instructions.

\subsection{Models and Fine-Tuning}

Three bases were selected:
\begin{itemize}
    \item \textbf{Qwen3-4B-Thinking-2507 (4B)}, representing a smaller thinking-oriented base~\citep{qwen2025qwen3};
    \item \textbf{DeepSeek-R1-0528-Qwen3-8B (8B)}, a reasoning-distilled model~\citep{deepseek2025r1};
    \item \textbf{Qwen3-8B (8B)}, the strongest standard base in the study.
\end{itemize}

Each base was fine-tuned on both data conditions, yielding six variants (Table~\ref{tab:experimental_design}).

\begin{table}[!t]
\centering
\caption{Original fine-tuning design. The two data conditions are token-matched to within 2\%.}
\label{tab:experimental_design}
\begin{tabular}{lccc}
\toprule
Training data & Qwen3-4B & DS-R1-8B & Qwen3-8B \\
\midrule
Dart + synthetic Dart & v1 & v3 & v5 \\
Dart + Swift & v2 & v4 & v6 \\
\bottomrule
\end{tabular}
\end{table}

The submitted run report describes LoRA with DoRA~\citep{liu2024dora}, attention and feed-forward targets, rank $r=32$, $\alpha=32$, dropout 0.09, Paged AdamW (8-bit), learning rate $2\times10^{-5}$, batch size 1, gradient accumulation 4, and label smoothing 0.1. The public candidate scripts instead contain multiple incompatible configurations (including rank 16 or 32, dropout 0.07 or 0.08, and learning rate $2\times10^{-4}$), and no immutable run ledger maps them to v1--v6. Consequently, these settings are the submitted report, not independently verified run metadata. Each variant has one archived training outcome; training-seed robustness is unmeasured. Training used one NVIDIA H200 GPU and was reported to require approximately 1.5 hours for 4B models and 3 hours for 8B models.

\subsection{Original Evaluation Protocols}

\subsubsection{\codebleu.}
The custom Dart \codebleu implementation is evaluated on CB-73 with five generations per function, temperature 0.2, top-$p=0.99$, and beam size 1. It combines n-gram, weighted n-gram, AST node-type overlap, and an identifier-level data-flow heuristic. It is a static similarity measure and does not establish behavioral equivalence.

\subsubsection{\compileatk.}
The original compilation study uses CP-126 with five attempts per function. A candidate is concatenated with the real test file and run with the Dart toolchain. A static front-end diagnostic counts as a compilation failure; runtime exceptions and failed assertions count as compiled but not passing. Because the harness calls a named function with typed arguments and consumes a typed return value, \compileatk substantially measures interface conformance and type consistency.

\subsubsection{\passatk.}
The historical functional study uses HE-154 with $n=10$ candidates per task and the unbiased estimator of \citet{chen2021evaluating}. Those generations expose the semantic HumanEval-Dart signature and therefore evaluate metadata-assisted reconstruction. The legacy scorer hard-coded task 160 (\texttt{doAlgebra}) to $n=10,c=0$ without generation or execution. Later evaluators remove this hook, but the original missing outcome cannot be recovered. We retain the historical denominator and label its scores as legacy. A genuine outcome could increase either arm's mean by up to $100/154=0.65$ percentage points and change a paired difference by up to $\pm0.65$ points; a shared forced zero does not guarantee an unbiased contrast. Excluding task 160 and worst-case outcome sensitivity checks are reported in Appendix~\ref{app:benchmark_audit}. Sections~\ref{sec:benchmark_audit} and~\ref{sec:construct_validity} add audited denominators and metadata controls.

\subsection{Common-Candidate Metric Audit}
\label{sec:aligned_metric_method}

The common-candidate audit measures metric association directly by scoring identical outputs for static similarity, compilation, and execution. It uses a separate graph-prefix study with test-hint fields removed. Matched Qwen3-8B no-edge systems were independently trained with \texttt{GRAPH\_SEED} 42, 43, and 44, producing three different checkpoint SHA-256 identities. Each run contains 154 tasks and exactly ten candidates per task. Every candidate is scored by the same Dart extractor for \codebleu, the same front end for compilation, and the same JIT-and-test path for execution. We report HE-154 and the strict-clean HE-141 projection. This is independent training within the follow-up; the original CB-73, CP-126, and HE-154 panels have different candidates, and the historical v1--v6 pools are unavailable for aligned rescoring.

For each task, the audit records best-of-10 \codebleu and the unbiased \compileatk and \passatk estimators. We compute task-level Spearman association at matched $k$, 95\% task-bootstrap intervals, the area under the ROC curve for best \codebleu discriminating tasks with at least one passing candidate, and candidate-level $P(\mathrm{pass}\mid\mathrm{compile})$. The enclosing 38-run provenance report is marked \texttt{FAIL} because two CLAP runs record an encoder revision different from the expected revision; the selected three no-edge runs use the expected revision. Candidate-level replay of the archived single-execution labels is canonical for Table~\ref{tab:aligned_metrics}. The HE-154 architectural summary separately applies the documented one-candidate stability overlay (Section~\ref{sec:metric_finding}); aggregate/replay discrepancies are documented separately in Appendix~\ref{app:architecture}. The immutable v5 deposit archives the aligned per-candidate \codebleu/compile/pass score records, evaluation configurations, provenance sidecars, analysis script and output, and hash receipts for the policy-omitted raw prediction payloads.

\subsection{Benchmark Audit}
\label{sec:benchmark_audit}

Every supplied HE reference was compiled and executed against its own harness under the pinned toolchain. This produces:
\begin{itemize}
\item 151/154 references that compile;
\item 142/154 references that pass;
\item 12 reference-uncertified tasks;
\item four independent contract-defective tasks (\#121, \#127, \#153, and \#161);
\item three tasks in both defect populations, leaving 141 tasks that meet both certification criteria.
\end{itemize}

The reference execution is a post-hoc certification of the retained 154-row JSONL. The repository begins with that finished file; the original translation conversation, exact prompt, execution transcript, reject log, and validation checklist are missing. The later author account records the reported construction procedure, not per-task validation evidence. Appendix~\ref{app:benchmark_audit} enumerates all ten omitted IDs and compares them with MultiPL-E's Dart task set. Exact local exclusion reasons remain unverified and are kept separate from compatible upstream explanations.

The metadata-ablation arms were rescored on HE-154, HE-150, HE-142, and HE-141. The historical nine-model fine-tuning matrix remains on HE-154; a complete clean-subset rescore is unavailable for that matrix.

The HE tasks were also checked against 2,496 training rows behind the metadata-ablation checkpoint using exact source, identifier-normalized source, exact assembly, function-name, and 7-gram near-duplicate comparisons at Jaccard $\geq0.8$. No overlap was detected under those five criteria. This is evidence against the checked forms of training overlap, not proof that all semantic duplicates or pretraining exposure are absent.

\subsection{MF-175 Representation and Interventions}

Each eligible MF function was compiled to a source-only AOT ELF with Dart 3.12.2 and represented as \textbf{F2}: the target function's disassembled instruction stream, a serialized basic-block control-flow graph, and literals recovered from object-pool references, emitted under a fixed prompt grammar. The replicated analyses remove the one acceptance-test overlap and use MF-174.

The channel interventions are:
\begin{itemize}
\item \textbf{typed contract}: replace the opaque signature with \fnz and declared parameter/return types;
\item \textbf{constants removed}: delete recovered string and numeric literals while preserving the structural payload;
\item \textbf{structure substituted}: assign another task's assembly and CFG while retaining the original task's constants;
\item \textbf{typed contract, no binary}: preserve the typed contract but replace the complete binary payload with the empty string.
\end{itemize}
The final arm is a gold-derived, out-of-distribution oracle rather than a deployable system.

\subsection{Statistical Procedures and Chronology}

The historical \codebleu comparisons use paired Wilcoxon tests, paired $t$-tests, effect sizes, and win/loss/tie counts. We treat these as exploratory and apply Holm adjustment to all eight available Wilcoxon contrasts. For functional model comparisons, each task contributes its unbiased \passatk estimate at $k\in\{1,5,10\}$. We use 100,000 paired task bootstrap resamples for 95\% intervals and a two-sided paired label-swap randomization test (one million draws) for the mean task-level difference. This test assumes that the paired outcomes are exchangeable under within-task label swaps under the null; treatment assignment was not randomized. Holm correction is applied jointly to the 18 variant--base endpoints. These tests estimate task-level differences conditional on the archived training outcomes. McNemar's exact test provides a secondary analysis of \emph{coverage@10}, the indicator $c>0$ among ten candidates. It uses each task as its own control, discards concordant pairs, and tests whether discordant wins and losses are balanced under a binomial($n_{\mathrm{disc}}$, 0.5) null. Its endpoint is any-success coverage, not \passat{1}.

For the MF repeated-seed analyses, arms share the same task set and seed blocks. We use paired sign flips over blocks, assuming within-block arm-label exchangeability under the null. With five blocks the smallest attainable two-sided $p$ is $2/2^5=0.0625$; with six it is $2/2^6=0.03125$. After the first five blocks were inspected, one additional seed was prospectively sealed for a single typed-contract-versus-baseline \passat{10} contrast and a hard stop. Because the extension was selected after initial results were known and includes a diagnostic seed, the six-block result is treated as supportive rather than as an independent preregistered replication.

We report 10,000-iteration task bootstrap intervals where available, block bootstraps for seed-level magnitude, Wilson intervals for rates, and the minimum attainable exact $p$ when discreteness limits inference. Non-significance is never interpreted as equivalence.

\subsection{Result Lineage}

All result files are classified by prompt schema, corpus hash, checkpoint identity, scorer revision, and candidate-level replay status. Historical pools affected by prompt leakage are retained as audit records but excluded from claims. The original v1--v6 archive has additional limits: task 160 was forced to zero, the aggregate Dart row count and reported training count differ, candidate training scripts conflict, and v3/v4 orphan summaries disagree with the canonical per-task files. We use the per-task files where available, do not call the orphan summaries repeated runs, and restrict claims to the archived configurations. Appendix~\ref{app:claim_ledger} provides the evidence and artifact map.

\section{Results}
\label{sec:results}

\noindent\textbf{Historical-contract note.} The fine-tuning matrix evaluates metadata-assisted reconstruction with semantic HumanEval-Dart signatures exposed. We recompute its aggregate results and paired inference from the archived per-task files; no new generations were run for the historical adapter comparisons. The benchmark audit and metadata controls in Sections~\ref{sec:benchmark_audit} and~\ref{sec:construct_validity}, together with the common-candidate audit below, are separately identified follow-up studies.

This section reports four findings: direct metric association and aggregate divergence, training-mixture contrasts, configuration-dependent adaptation outcomes, and the positive but inconclusive v1 estimate. The historical matrix uses per-task summaries; the separate aligned audit provides candidate-level replay across three training seeds.

\subsection{Complete Evaluation Matrix}

Table~\ref{tab:complete_results} presents the historical nine-model matrix plus one external API reference. Because the metrics use different task panels and candidate pools, columns must not be compared as if they were aligned observations.

\begin{table*}[!t]
\centering
\caption{Complete evaluation results. \textsc{CodeBLEU} is best-of-$k$ on 73 Dart functions; \texttt{compile@k} uses 126 functions (630 attempts); and \texttt{pass@k} uses 154 HumanEval-Dart tasks ($n{=}10$ per task). The Qwen3-Max row is a metadata-assisted reference point. \textbf{Bold} indicates improvement over the respective base; \textit{italic} indicates regression.}
\label{tab:complete_results}
\scriptsize
\setlength{\tabcolsep}{3pt}
\begin{tabularx}{\textwidth}{Y l l c c c c c c}
\toprule
Model & Base & Training & \textsc{CodeBLEU} & \texttt{compile@1} & \texttt{compile@5} & \texttt{pass@1} & \texttt{pass@5} & \texttt{pass@10} \\
\midrule
Qwen3-Max (external reference) & API & --- & 0.663 & --- & --- & 18.38\% & 25.29\% & 27.92\% \\
\midrule
\multicolumn{9}{l}{\textit{Qwen3-8B family}} \\
Qwen3-8B (base) & 8B & --- & 0.767 & 39.5\% & 60.3\% & 6.36\% & 14.89\% & 18.18\% \\
decompiler-v5 & 8B & Dart+Synth & \textit{0.746} & \textit{35.9\%} & \textbf{61.1\%} & \textit{0.71\%} & \textit{2.62\%} & \textit{3.90\%} \\
decompiler-v6 & 8B & Dart+Swift & \textit{0.735} & \textit{29.5\%} & \textit{55.6\%} & \textit{1.10\%} & \textit{4.18\%} & \textit{5.84\%} \\
\midrule
\multicolumn{9}{l}{\textit{DeepSeek-R1-Qwen3-8B family}} \\
DeepSeek-R1-8B (base) & 8B & --- & 0.671 & 25.1\% & 62.7\% & 3.90\% & 11.04\% & 14.29\% \\
decompiler-v3 & 8B & Dart+Synth & \textbf{0.722} & \textbf{41.1\%} & \textbf{82.5\%} & \textit{3.12\%} & \textit{9.02\%} & \textit{12.34\%} \\
decompiler-v4 & 8B & Dart+Swift & \textbf{0.692} & \textbf{32.7\%} & \textbf{74.6\%} & \textit{2.40\%} & \textit{6.17\%} & \textit{7.79\%} \\
\midrule
\multicolumn{9}{l}{\textit{Qwen3-4B family}} \\
Qwen3-4B (base) & 4B & --- & 0.658 & 35.9\% & 71.4\% & 3.25\% & 9.30\% & 12.34\% \\
decompiler-v1 & 4B & Dart+Synth & \textbf{0.723} & \textit{31.6\%} & \textit{69.8\%} & \textbf{3.96\%} & \textbf{11.97\%} & \textbf{16.23\%} \\
decompiler-v2 & 4B & Dart+Swift & \textbf{0.679} & \textit{14.3\%} & \textit{39.7\%} & \textit{1.30\%} & \textit{3.81\%} & \textit{5.19\%} \\
\bottomrule
\end{tabularx}
\end{table*}

The Qwen3-8B base has the highest historical \passat{1} among the three local bases, and both of its adapters regress after Holm correction across the 18 \passatk endpoints. The other four adapter comparisons are inconclusive. Some static and compilation columns rise while functional point estimates fall on the separate panel, motivating the aligned metric audit below. Training-mixture contrasts also vary by base configuration. The external Qwen3-Max row provides a descriptive reference, not a parameter-scaling comparison.

\subsection{Statistical Validation}

Table~\ref{tab:codebleu_stats} presents paired statistical tests on the 73-sample \textsc{CodeBLEU} evaluations.

\begin{table}[!t]
\centering
\caption{Historical paired \textsc{CodeBLEU} contrasts on CB-73. Holm adjustment covers all eight available Wilcoxon contrasts. Win/loss/tie uses a 0.01 margin.}
\label{tab:codebleu_stats}
\begin{tabular}{lccccc}
\toprule
Comparison & $\Delta$ & Raw $p$ & Holm $p$ & $d$ & W/L/T \\
\midrule
v1 vs. 4B base & $+.065$ & .033 & .155 & $+.32$ & 41/25/7 \\
v2 vs. 4B base & $+.021$ & .252 & .503 & $+.09$ & 22/43/8 \\
v3 vs. R1-8B & $+.051$ & .001 & .006 & $+.32$ & 46/18/9 \\
v4 vs. R1-8B & $+.020$ & .067 & .202 & $+.11$ & 41/23/9 \\
v1 vs. v2 & $+.045$ & $<.001$ & .003 & $+.33$ & 50/16/7 \\
v3 vs. v4 & $+.031$ & .026 & .155 & $+.18$ & 40/23/10 \\
v5 vs. 8B base & $-.021$ & .580 & .580 & $-.18$ & 38/22/13 \\
v6 vs. 8B base & $-.032$ & .028 & .155 & $-.38$ & 23/32/18 \\
\bottomrule
\end{tabular}
\end{table}

The v3--R1 and v1--v2 \codebleu contrasts remain detectable after Holm adjustment; the other six do not. These exploratory CB-73 comparisons describe static similarity under the evaluated configurations.

\subsubsection{Task-Level \texttt{pass@k} Significance Testing}

\begin{figure}[!t]
\centering
\includegraphics[width=\linewidth]{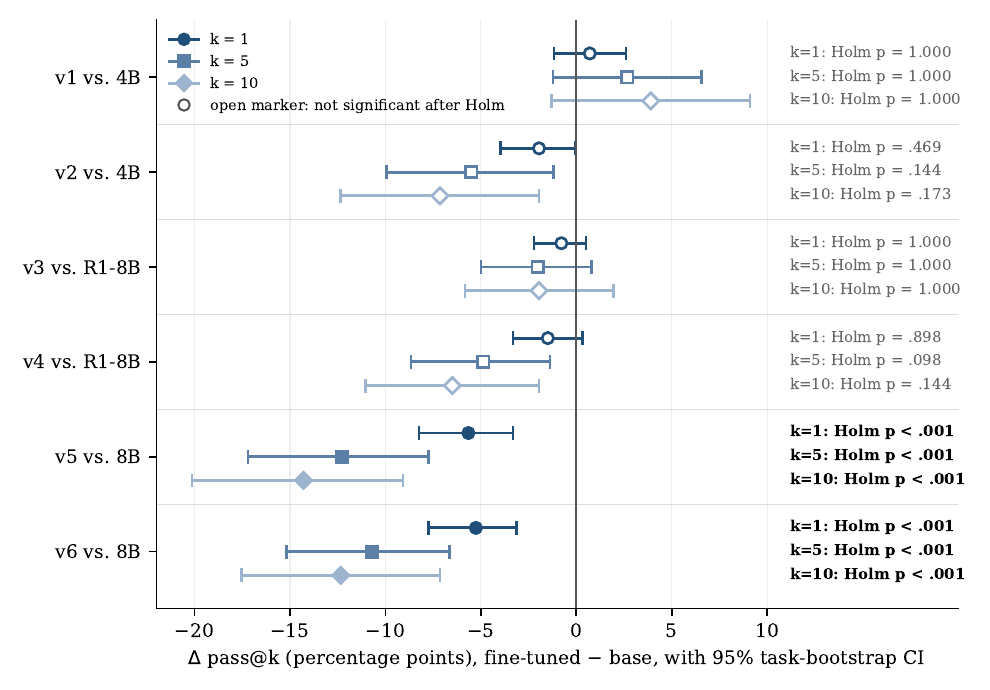}
\Description{Forest plot of six fine-tuned minus base comparisons, each at pass@1, pass@5, and pass@10. Pointwise intervals cross zero for several contrasts. Only the v5 and v6 regressions retain significance after the 18-test Holm correction.}
\caption{\passatk differences (fine-tuned minus base) for the six historical comparisons in Table~\ref{tab:passk_tests_corrected}, with pointwise 100,000-resample task-bootstrap 95\% intervals at $k=1,5,10$. Filled markers denote significance after Holm correction over 18 tests; open markers do not. Holm adjusts $p$-values, not the plotted estimates or intervals. Only v5 and v6 remain significant at every $k$. Values are taken from that table.}
\label{fig:passk_forest}
\end{figure}

The primary tests align inference with \passatk at $k\in\{1,5,10\}$ using each task's unbiased estimator. Table~\ref{tab:passk_tests_corrected} reports the paired mean difference, a 100,000-resample task-bootstrap interval, a two-sided paired label-swap randomization $p$-value, and Holm adjustment over all 18 endpoints. Fig.~\ref{fig:passk_forest} shows the differences and pointwise intervals; Holm adjustment applies to the $p$-values, not the estimates or intervals. Exact integer-arithmetic replay in the artifact confirms the Monte Carlo decisions. This corrects the original McNemar labeling: that test used whether \emph{any} of ten candidates passed ($c>0$), which is coverage@10, not \passat{1}. All primary inference here is conditional on the archived training outcomes.

\begin{table*}[!t]
\centering
\caption{Corrected task-level inference on HE-154. Differences are fine-tuned minus base in percentage points. $p$ uses one million paired label swaps; Holm adjustment covers all $6\times3=18$ tests.}
\label{tab:passk_tests_corrected}
\scriptsize
\setlength{\tabcolsep}{4pt}
\begin{tabular}{lcrlrr}
\toprule
Comparison & $k$ & $\Delta$ (pp) & 95\% CI (pp) & Raw $p$ & Holm $p$ \\
\midrule
v1 vs. 4B & 1 & $+0.71$ & $[-1.17,+2.60]$ & .5065 & 1.000 \\
             & 5 & $+2.67$ & $[-1.21,+6.57]$ & .1846 & 1.000 \\
             & 10 & $+3.90$ & $[-1.30,+9.09]$ & .2100 & 1.000 \\
\midrule
v2 vs. 4B & 1 & $-1.95$ & $[-3.96,-0.06]$ & .0587 & .469 \\
             & 5 & $-5.50$ & $[-9.94,-1.20]$ & .0134 & .144 \\
             & 10 & $-7.14$ & $[-12.34,-1.95]$ & .0192 & .173 \\
\midrule
v3 vs. R1-8B & 1 & $-0.78$ & $[-2.21,+0.52]$ & .3147 & 1.000 \\
             & 5 & $-2.02$ & $[-5.00,+0.81]$ & .1769 & 1.000 \\
             & 10 & $-1.95$ & $[-5.84,+1.95]$ & .5078 & 1.000 \\
\midrule
v4 vs. R1-8B & 1 & $-1.49$ & $[-3.31,+0.32]$ & .1282 & .898 \\
             & 5 & $-4.87$ & $[-8.65,-1.37]$ & .0082 & .098 \\
             & 10 & $-6.49$ & $[-11.04,-1.95]$ & .0131 & .144 \\
\midrule
v5 vs. 8B & 1 & $-5.65$ & $[-8.25,-3.31]$ & .000001 & .000018 \\
             & 5 & $-12.27$ & $[-17.21,-7.73]$ & .000001 & .000018 \\
             & 10 & $-14.29$ & $[-20.13,-9.09]$ & .000001 & .000018 \\
\midrule
v6 vs. 8B & 1 & $-5.26$ & $[-7.73,-3.12]$ & .000001 & .000018 \\
             & 5 & $-10.70$ & $[-15.17,-6.64]$ & .000001 & .000018 \\
             & 10 & $-12.34$ & $[-17.53,-7.14]$ & .000009 & .000117 \\
\bottomrule
\end{tabular}
\end{table*}

The corrected 18-endpoint Holm family identifies v5 and v6 regressions at every $k$ and leaves v1--v4 inconclusive. In the secondary coverage@10 analysis, gained/lost tasks are 11/5, 4/15, 3/6, 2/12, 0/22, and 0/19 for v1--v6 against their respective bases; exact McNemar $p$-values are .210, .019, .508, .013, $4.8\times10^{-7}$, and $3.8\times10^{-6}$. The two additional nominal findings do not survive correction across the six coverage contrasts.

A second archived inference draw retains the direction of the Qwen3-8B/v5 regression. Base \passat{1}/\passat{5}/\passat{10} changes from 6.36/14.89/18.18\% to 6.10/14.14/17.53\%, while v5 changes from 0.71/2.62/3.90\% to 1.10/4.19/6.49\%. Per-task pass counts agree exactly across draws on 130/154 base tasks and 143/154 v5 tasks. On the second draw, v5--base differences are $-5.00$, $-9.95$, and $-11.04$ pp at $k=1,5,10$ (paired randomization $p=.000005$, $.000002$, and $.000078$). These files lack full seed and candidate provenance and represent inference repetition, not repeated historical training. Independent training-seed variation is measured in the separate graph-prefix follow-up below.

\subsection{Candidate-Level Metric Association Under the Metadata Intervention}
\label{sec:metric_by_contract}

The semantic-metadata intervention provides 1,540 candidates per arm from one frozen checkpoint and paired seed-42 run. Each output receives \codebleu, compilation, and execution scores, enabling direct candidate- and task-level association analysis. Table~\ref{tab:metadata_ablation} in Section~\ref{sec:metadata_intervention} reports the corresponding aggregate rates.

\begin{table}[!t]
\centering
\caption{Candidate- and task-level metric association on the semantic-metadata intervention (one checkpoint, paired seed-42 run, HE-154, $n=1{,}540$ candidates per arm; same records as Table~\ref{tab:metadata_ablation}). $\rho$ values are Spearman rank correlations.}
\label{tab:metric_by_contract}
\small
\setlength{\tabcolsep}{3.5pt}
\begin{tabular}{lrr}
\toprule
 & Semantic metadata & Typed opaque \fnz \\
\midrule
\multicolumn{3}{l}{\emph{Candidate level} ($n=1{,}540$)} \\
Compiled & 77.08\% & 84.61\% \\
Passed tests & 17.66\% & 0.71\% \\
$P(\text{pass}\mid\text{compiled})$ & 22.9\% & 0.8\% \\
$\rho(\codebleu,\text{pass})$ & $+0.29$ & $-0.00$ \\
$\rho(\codebleu,\text{compile})$ & $+0.14$ & $+0.12$ \\
$\rho(\text{compile},\text{pass})$ & $+0.25$ & $+0.04$ \\
\midrule
\multicolumn{3}{l}{\emph{Task level} ($n=154$)} \\
$\rho(\text{mean }\codebleu,\passat{1})$ & $+0.30$ & $-0.21$ \\
$\rho(\text{compile count},\text{ pass count})$ & $+0.22$ & $+0.06$ \\
$P(\text{any-pass}\mid\text{all ten compile})$ & 38.6\% & 2.4\% \\
\bottomrule
\end{tabular}
\end{table}

Metric association varies with the input contract in this run (Table~\ref{tab:metric_by_contract}). Under semantic metadata, the reported correlations are at most .30. Under typed opaque metadata, candidate-level associations are near zero or small (up to .12), while the \emph{task-level} correlation between mean \codebleu and \passat{1} is negative ($-.21$). These descriptive coefficients use distinct units: candidates clustered within tasks and task-level summaries from this single run.

\subsection{Finding 1: Direct Metric Association and Aggregate Divergence}
\label{sec:metric_finding}

The common-candidate audit finds modest \codebleu association and weak compilation association across three independently trained seeds. Table~\ref{tab:aligned_metrics} answers RQ4 on the strict-clean HE-141 subset for the same no-edge graph-prefix configuration; all entries in a row derive from the same ten candidates per task. It complements the single-run contract comparison in Table~\ref{tab:metric_by_contract}.

\begin{table*}[!t]
\centering
\caption{Aligned metric audit on HE-141 using archived single-execution outcome labels. $\rho$ is task-level Spearman correlation at $k=10$; the final column is candidate-level conditional success. These rows precede the documented stability overlay used in the HE-154 core summary.}
\label{tab:aligned_metrics}
\small
\begin{tabular}{crrrrrr}
\toprule
Seed & Best \codebleu & \compileat{10} & \passat{10} & $\rho$(CB, pass) & $\rho$(compile, pass) & $P(\mathrm{pass}\mid\mathrm{compile})$ \\
\midrule
42 & .677 & 98.58\% & 27.66\% & .218 & .074 & 23.32\% \\
43 & .672 & 98.58\% & 26.24\% & .246 & .072 & 20.98\% \\
44 & .670 & 98.58\% & 31.91\% & .240 & .082 & 22.8\% \\
\bottomrule
\end{tabular}
\end{table*}

Best \codebleu has modest positive association with functional coverage: bootstrap 95\% intervals for $\rho$(\codebleu, \passat{10}) are [.054,.374], [.086,.396], and [.074,.396], and corresponding AUCs for classifying coverage@10 are .641, .661, and .648. Compilation is a necessary gate but a weak discriminator in this pool. \compileat{10} is nearly saturated, its task-level correlation with \passat{10} is .072--.082, and 21.0--23.3\% of compiling candidates pass all tests. Associations at $k=1$ and $k=5$ have the same qualitative ordering: \codebleu correlations are .245--.278 and .230--.272, whereas matched compilation correlations are .132--.244 and .099--.136.

The historical v3 contrast illustrates the practical distinction between metrics: on separate panels, \codebleu rises from .671 to .722 and \compileat{1} from 25.1\% to 41.1\%, while \passat{1} moves from 3.90\% to 3.12\%. The aligned follow-up supplies direct same-candidate association and discrimination evidence: static similarity is modestly associated with task-level coverage, compilation filters invalid programs, and execution distinguishes the compiling candidates that satisfy the tests.

The three rows also measure training-seed variation: distinct checkpoint SHA-256 values at seeds 42, 43, and 44 document independent training with a matched prompt stream and configuration. The HE-154 core summary additionally applies the documented pass-stability overlay: at seed 44, candidate 4 for problem 158 (row 150) changes from passing to failing. The archive records repeated failure of this candidate's stochastic tie-breaking under the stability criterion; this is a historical correction, not a new execution in the present audit.

After that overlay, the no-edge configuration has $\passat{1}=.160\pm.010$, $\passat{10}=.260\pm.023$, and $\compileat{1}=.756\pm.007$ (mean $\pm$ sample SD across training seeds). Coverage counts are 39, 37, and 44 of 154, with 260, 231, and 250 passing candidates of 1,540. Without the overlay, the archived flags give 39, 37, and 45 covered tasks, $\passat{1}=.161\pm.010$, and $\passat{10}=.262\pm.027$; Table~\ref{tab:aligned_metrics} uses those archived single-execution labels. Two further matched graph-prefix variants were independently trained at the same three seeds; their stability-qualified core summaries appear in Appendix~\ref{app:architecture}. This repeated training applies to the follow-up, not the historical adapter matrix.

The enclosing 38-run provenance report remains \texttt{FAIL} because of two CLAP encoder-revision mismatches (Section~\ref{sec:aligned_metric_method}); the selected no-edge seeds use the expected revision.

\subsection{Finding 2: Observed Training-Mixture Contrasts}

Training-mixture contrasts vary across the three base configurations (Table~\ref{tab:mixture_descriptive}). At Qwen3-4B, the Dart--Swift variant is lower on all three historical panels; at DeepSeek-R1-Qwen3-8B the gap is smaller; at Qwen3-8B both adapters are far below the base in \passatk and their between-adapter difference is small.

\begin{table}[!t]
\centering
\caption{Historical Dart+Synthetic (D+S) and Dart+Swift (D+Sw) results. Metrics use separate panels; no causal language or scale effect is identified.}
\label{tab:mixture_descriptive}
\small
\begin{tabular}{llrrr}
\toprule
Base configuration & Mixture & \codebleu & \compileat{5} & \passat{1} \\
\midrule
Qwen3-4B & D+S & .723 & 69.8\% & 3.96\% \\
          & D+Sw & .679 & 39.7\% & 1.30\% \\
DeepSeek-R1-Qwen3-8B & D+S & .722 & 82.5\% & 3.12\% \\
          & D+Sw & .692 & 74.6\% & 2.40\% \\
Qwen3-8B & D+S & .746 & 61.1\% & 0.71\% \\
          & D+Sw & .735 & 55.6\% & 1.10\% \\
\bottomrule
\end{tabular}
\end{table}

For RQ2--RQ3, Dart+Synthetic has the higher historical point estimate in all six displayed columns across 4B and R1-8B, whereas both Qwen3-8B adapters show functional regression. These comparisons concern the evaluated pipelines: optimized Dart AOT and Swift \texttt{-O0} vary in language, compiler optimization, and corpus composition, while the bases also vary in architecture, pretraining, and reasoning/distillation. A factorial language-by-optimization comparison with matched data, together with a controlled single-family size sweep, would test the language and capacity explanations.

\subsection{Finding 3: Model-Configuration-Dependent Fine-Tuning Outcomes}

For the Dart+Synthetic variants, the corrected \passat{1} comparisons are summarized in Table~\ref{tab:ft_by_config}. The Qwen3-4B and DeepSeek-R1-Qwen3-8B estimates are inconclusive after multiplicity correction; Qwen3-8B shows a large regression. The same conclusion holds at $k=5$ and $k=10$ (Table~\ref{tab:passk_tests_corrected}).

\begin{table}[!t]
\centering
\caption{Dart+Synthetic adaptation by base configuration. Holm $p$ belongs to the 18-endpoint family and tests the task-level \passat{1} difference.}
\label{tab:ft_by_config}
\small
\begin{tabular}{lrrrl}
\toprule
Base configuration & Base & Adapter & $\Delta$ (pp) & Holm $p$ \\
\midrule
Qwen3-4B & 3.25\% & 3.96\% & $+0.71$ & 1.000 \\
DeepSeek-R1-Qwen3-8B & 3.90\% & 3.12\% & $-0.78$ & 1.000 \\
Qwen3-8B & 6.36\% & 0.71\% & $-5.65$ & $<.00003$ \\
\bottomrule
\end{tabular}
\end{table}

The outcome is model-configuration dependent in this experiment. Qwen3-4B-Thinking, DeepSeek-R1-Qwen3-8B, and Qwen3-8B differ in architecture or configuration, pretraining, reasoning/distillation, and baseline performance as well as parameter count. A controlled family sweep with repeated training and tuned hyperparameters would test a scale explanation; the present comparison does not isolate it.

\subsection{Finding 4: Decompiler-v1 Has a Positive but Inconclusive Point Estimate}

Decompiler-v1 is the only adapter with positive historical point estimates at all three $k$ values: $+0.71$ pp at \passat{1}, $+2.67$ pp at \passat{5}, and $+3.90$ pp at \passat{10}. All intervals cross zero and all Holm-adjusted $p$-values equal 1.000. Coverage@10 has 11 gains and 5 losses (unadjusted McNemar $p=.210$). The gained tasks have median 26 parsed instructions and the lost tasks median 116; the corresponding raw non-empty disassembly-line medians are 42 and 122. Reporting both quantities resolves the earlier line/instruction inconsistency. These identities come from one historical draw and are descriptive.

The CB-73 \codebleu estimate rises from .658 to .723, with an eight-contrast Holm-adjusted Wilcoxon $p_{\mathrm{Holm}}=.155$. Thus v1's positive static and functional point estimates remain inconclusive.

\subsection{Stratified Analysis: Fine-Tuning Effect by Instruction Count}

Table~\ref{tab:bin_stratified_revised} summarizes the historical draw in four post-hoc, round-number ranges ($<50$, 50--99, 100--199, and $\geq200$ parsed instructions). The ranges were chosen for readability, not as quantiles or through outcome-optimized threshold selection.

\begin{table}[!t]
\centering
\caption{Historical \passat{1} (\%) by parsed instruction count. Small cells and post-hoc bins preclude causal or threshold claims.}
\label{tab:bin_stratified_revised}
\small
\begin{tabular}{lrrrrrrr}
\toprule
Instructions & 4B & v1 & 8B & v5 & R1 & v3 & $n$ \\
\midrule
$<50$ & 2.1 & 9.4 & 9.4 & 0.6 & 5.0 & 4.4 & 34 \\
50--99 & 3.1 & 4.5 & 8.2 & 0.6 & 4.7 & 4.3 & 51 \\
100--199 & 5.0 & 1.1 & 4.4 & 1.1 & 3.5 & 2.0 & 54 \\
$\geq200$ & 0.0 & 0.0 & 0.0 & 0.0 & 0.0 & 0.0 & 15 \\
\bottomrule
\end{tabular}
\end{table}

The v1 point estimate is higher below 100 instructions and lower at 100--199; v5 is below its base in every range. No evaluated configuration records a passing candidate at $\geq200$ parsed instructions in this draw. This is a length association in the sampled task set, not a discovered capability threshold; instruction count also varies with control flow, source complexity, prompt truncation risk, and benchmark topic.

\subsection{\textsc{CodeBLEU} Distribution Shift}

The historical CB-73 distributions differ most in their lower tails: the 4B base and v1 have 10th percentiles .025 and .408, respectively; 13 versus 3 outputs score below .3; and their medians are .722 and .782. The 25th percentile moves in the opposite direction (.592 to .547), and the Holm-adjusted paired contrast is not significant. These summaries describe where static scores change within CB-73; functional task gains and losses are measured separately on HE-154.

\subsection{Compilation Coverage and Conditional Static Quality}

On CP-126, the DeepSeek-R1-Qwen3-8B base has 79 compiling outputs with mean compiled-only \codebleu .585, while v3 has 104 with mean .587. Qwen3-8B has 76 compiling outputs with mean .666, versus 77/.632 for v5 and 70/.608 for v6. The R1/v3 pair therefore combines broader compilation coverage with similar static quality among compiling outputs. This is a conditional summary: filtering on compilation changes the compared sample, and CP-126 is separate from HE-154.

The historical solved sets also differ across configurations. Thirty-seven HE-154 tasks are covered by at least one of the nine local configurations, seven by exactly one, and none by all nine. These one-draw counts describe observed overlap; Section~\ref{sec:sampling_reliability} examines how task identities change under repeated decoding.

\section{Error and Complexity Analysis}
\label{sec:error}

Across the nine local configurations, 37/154 tasks are covered at least once and 117 are not observed covered. Here ``covered'' means that at least one candidate in the historical generation pool passed the available harness. We use those observed labels to describe task length and code features, rather than intrinsic solvability.

\begin{table}[!t]
\centering
\caption{Observed historical coverage by parsed instruction count. Bins are post hoc and use one task-level draw.}
\label{tab:length_coverage_revised}
\small
\begin{tabular}{lrrr}
\toprule
Instructions & Tasks & Covered & Rate \\
\midrule
$<50$ & 34 & 14 & 41\% \\
50--99 & 51 & 13 & 25\% \\
100--199 & 54 & 10 & 19\% \\
200--499 & 14 & 0 & 0\% \\
$\geq500$ & 1 & 0 & 0\% \\
\bottomrule
\end{tabular}
\end{table}

Covered tasks have fewer parsed instructions on average than uncovered tasks (72.2 versus 134.1; one-sided Mann--Whitney $p<.001$). No adjacent-bin two-sided Fisher comparison is significant; the $\geq200$ boundary gives $p=.103$ against 100--199. Thus the evidence supports a length association, not a sharp transition. Instruction count covaries with task content, compiler transformations, runtime calls, and prompt budget. Earlier cyclomatic-complexity and nesting-depth results are omitted because their pipelines cannot be re-derived from the frozen artifact.

For v1 versus its base, the 11 coverage gains have median 26 parsed instructions and the 5 losses median 116 (42 and 122 raw non-empty disassembly lines, respectively). These recomputed, explicitly unit-labeled values supersede both inconsistent pairs in the submitted version (27/110 and 44/124), neither of which is reproducible from the frozen artifact (Section~\ref{sec:assembly_generation}). List-using tasks are more frequent among the uncovered group, while Map-using tasks appear only there. These unadjusted patterns vary with task topic and length; repeated task-level frequencies would be needed to assess persistent gains or failure classes.

\section{Construct Validity of Functional Evaluation}
\label{sec:construct_validity}

The construct-validity study measures how functional success changes when semantic metadata is removed and when graph inputs are perturbed. These interventions address RQ5 by separating dependence on the supplied contract from evidence of instruction-derived recovery.

\subsection{Semantic-Metadata Intervention}
\label{sec:metadata_intervention}

The paired semantic-to-typed-opaque intervention freezes one checkpoint, the HE tasks, the scoring stack, and the Dart toolchain while varying the semantic-metadata bundle. The contract changes from semantic schema \nolinkurl{antigravity-v2-no-test-hints} to typed-opaque schema \nolinkurl{antigravity-v3-matched-function-contract}. A distinct opaque name-only condition uses the latter schema and is reported alongside that pair. The three contracts are:
\begin{itemize}
\item \textbf{semantic}: the original HumanEval-Dart name and parameter names, together with symbols reachable through disassembly and prompt metadata;
\item \textbf{typed opaque}: the same parameter and return types and arity, but the function name becomes \fnz, parameter names become positional, helper and disassembly symbols are neutralized, and the source path is removed;
\item \textbf{opaque name only}: the same binary soft-prefix with neutral \fnz as the only exposed function-contract field, without declared types or arity (the artifact records \texttt{public\_signature\_mode=name\_only}).
\end{itemize}

The intervention removes a metadata \emph{bundle}. Across all 154 tasks it replaces the function name, 200 parameter-name slots, disassembly headers and branch-target annotations, 15 helper symbols, and the source path. After masking symbol annotations, the opcode, operand, address, type, and arity fields remain identical in 154/154 tasks.

\begin{table}[!t]
\centering
\caption{Semantic-metadata intervention on one frozen checkpoint and paired seed-42 run. Types, arity, and instruction content are fixed between the first two rows. \texttt{compile@1} is the candidate-level compilation fraction; with ten candidates per task it also equals the mean per-task compilation fraction, not any-compile task coverage.}
\label{tab:metadata_ablation}
\begin{tabular}{lrrrr}
\toprule
Contract shown & \passat{1} & \passat{10} & compile@1 & Compiling candidates \\
\midrule
Semantic metadata & 17.66\% & 27.27\% & 77.08\% & 1,187/1,540 \\
Typed opaque \fnz & 0.71\% & 4.55\% & 84.61\% & 1,303/1,540 \\
Opaque name only + binary & 0.00\% & 0.00\% & 3.18\% & 49/1,540 \\
\bottomrule
\end{tabular}
\end{table}

Removing the metadata bundle reduces functional coverage from 42 tasks to 7 at \passat{10}, with 35 losses and 0 gains (exact two-sided $p\approx6\times10^{-11}$). Compilation increases: the typed opaque arm emits \fnz with the requested arity in 1,537/1,540 candidates, of which 1,303 compile against the tests. The distinct opaque name-only arm solves no task and compiles 49/1,540 candidates. The paired 35-loss/0-gain test concerns semantic versus typed opaque metadata; the seven-to-zero difference does not isolate removal of types or arity.

Lower functional success alongside higher compilation identifies a large within-run sensitivity to the specification bundle. The single-seed and off-distribution limits of this contrast are detailed in Section~\ref{sec:threats}.

\subsection{Sensitivity to Benchmark Defects}

\begin{table}[!t]
\centering
\caption{The paired metadata discordance is unchanged on every audited HumanEval-Dart denominator.}
\label{tab:metadata_denoms}
\begin{tabular}{lrrrrr}
\toprule
Denominator & $n$ & Semantic solved & Opaque solved & Losses & Gains \\
\midrule
Historical full set & 154 & 42 (27.27\%) & 7 (4.55\%) & 35 & 0 \\
Contract-valid & 150 & 42 (28.00\%) & 7 (4.67\%) & 35 & 0 \\
Reference-certified & 142 & 42 (29.58\%) & 7 (4.93\%) & 35 & 0 \\
Strict clean & 141 & 42 (29.79\%) & 7 (4.96\%) & 35 & 0 \\
\bottomrule
\end{tabular}
\end{table}

Every excluded task lies among the 112 solved by neither arm, so the metadata discordance is unchanged across the audited denominators. The five-criterion overlap audit also detects no HE row in the checked fine-tuning corpus. This supports the reported within-run contrast while leaving benchmark recall and inference from descriptive metadata as distinct possible explanations.

\subsection{Signature-Only Comparator}

A complementary study compares three systems that consume binary representations against the pretrained base model without task-specific fine-tuning, prompted with the semantic signature alone. The signature-only model receives no adapter and no binary, yet reaches \passat{10} of 36/154.

\begin{table}[!t]
\centering
\caption{Binary-conditioned arms against a pretrained, unadapted signature-only baseline. Intervals are pointwise paired task bootstraps; $p$ is unadjusted exact McNemar.}
\label{tab:signature_only}
\begin{tabular}{lrrr}
\toprule
Arm versus signature-only & $\Delta\passat{10}$ & 95\% CI & $p$ \\
\midrule
Fine-tuned on raw assembly & $-0.0195$ & $[-0.065,+0.026]$ & 0.549 \\
Graph prefix, no edges & $+0.0195$ & $[-0.020,+0.058]$ & 0.549 \\
Graph prefix + CFG & $+0.0260$ & $[-0.013,+0.071]$ & 0.388 \\
\bottomrule
\end{tabular}
\end{table}

The signature-only score provides a descriptive reference for performance under a semantic prompt without binary input. The between-system contrasts are inconclusive even before multiplicity adjustment; adaptation differs, so they establish neither an isolated binary contribution nor equivalence.

The opaque name-only arm retains the binary soft-prefix and exposes only neutral \fnz as its contract; the no-binary comparator exposes the full semantic signature. A matched semantic-name-only/no-binary arm was not run. Section~\ref{sec:assembly_ablation} tests the graph channel on two HE-154 input views. In those six runs, the decoder's \texttt{Assembly:} field contains the fixed line \texttt{; (assembly provided via graph channel)}, and instructions enter through the graph encoder. The raw-assembly systems use a different input path, and no raw-text shuffle was run for historical v1--v6. Appendix~\ref{app:architecture} reports the separate architecture diagnostics.

\subsection{Instruction-Shuffle and Cross-Task Graph Ablation}
\label{sec:assembly_ablation}

The graph ablations test dependence on the binary inputs consumed by the encoder. We use the frozen checkpoint \texttt{qwen3-8b-base\allowbreak{}\_lora\_enc\allowbreak{}\_dec\_r64\allowbreak{}\_5e6\_gcb\allowbreak{}\_a128\_graphv2\allowbreak{}\_clean\_s42\allowbreak{}\_prefix\_no\allowbreak{}\_gine\_regions16} (SHA-256 \texttt{e8e87260\ldots c347fc}), seed 42, and ten candidates per task. Each intervention has a newly generated, same-checkpoint baseline within its input view. These single-seed controls were added in response to review and were not preregistered.

The two files are \emph{views of the same 154 HumanEval tasks}, not independent datasets: \texttt{grpo\_data\_graphv2.jsonl} retains semantic metadata, while \texttt{grpo\_data\_graphv2\allowbreak{}\_sigscrub\_v3\allowbreak{}\_opaque\_nameonly\allowbreak{}\_public.jsonl} supplies the opaque name-only contract. Their SHA-256 prefixes are \texttt{8453876a} and \texttt{75c9a6c6}; the artifact records full hashes, candidate files, and score sidecars. Generation uses pinned encoder and decoder revisions, batch size 10, at most 768 new tokens, and a 2,048-token decoder prompt limit. Scoring uses Dart 3.11.5, within-view matched evaluator code, and a 30-second test timeout. The semantic view was scored on Windows and the opaque view on Linux; no between-view contrast is treated as a matched graph-ablation test.

The retained mappings verify that the interventions reach the encoder. All six sidecars record \texttt{assembly\_input\_ablation=none}; the graph modes are \texttt{none}, \texttt{shuffle\_instructions}, and \texttt{cyclic\_shift}. The hash-matched inference source calls \texttt{build\_blocks(graph\_rows[idx],\ldots)} on the altered records. Thus these runs change consumed graph input rather than the unused literal-assembly field. The artifact replays graph mappings and decoder-prompt hashes and retains both prompt-builder snapshots: schema v2 for the semantic view and v3's matched top-level-function guidance for the opaque view. The scorer snapshots differ in code extraction and trusted harness imports but are fixed within each three-arm comparison. Cross-view differences therefore involve more than one factor. Sidecars report non-strict checkpoint loading with 602 missing and no unexpected keys. Because the missing-key names were not logged, the checkpoint hash does not attest every loaded parameter; the count alone does not identify a loading fault.

Compilation is the archived JIT-harness classification, not a separate successful \texttt{dart compile} invocation: front-end diagnostics count as non-compilation, while runtime failures and timeouts count as compiled-but-failing. A timeout by itself does not prove front-end acceptance. The scorer also adds a random completion marker to reject early exits; the marker is not retained separately, so its instrumented-source hash is not byte-replayable. The archive verifies raw candidate hashes and provides the evaluator-only tests for execution replay, but does not claim that statistical replay re-executes Dart or reproduces the missing random markers.

\textbf{Instruction shuffle.} Within each task, a seeded derangement permutes the flattened CFG instruction list and fills the original block slots. All 18,359 source indices move, across 154/154 tasks; instructions are not exchanged between tasks. Repeated strings mean that 116 slots in the semantic view and 149 in the opaque view retain identical text despite a different source index. The per-task instruction multiset, block sizes, and stored CFG/edge topology are preserved. Instruction-derived within-block data flow is rebuilt downstream, so this is not a topology-only intervention or a guarantee that all derived graph features stay fixed.

\textbf{Cross-task substitution.} A cyclic offset of 43 replaces each target's \texttt{cfg}, \texttt{edges}, \texttt{integrity}, and \texttt{graph\_v2} fields with another task's payload. All 154 donor assignments are non-self. Target prompts and scoring tests are unchanged. This removes target--graph correspondence but is not length-matched and does not preserve the target topology.

\begin{table*}[!t]
\centering
\caption{Functional and compilation outcomes, seed 42, semantic HE-154 view. Within each condition all six rates use the same ten candidates per task and the same Dart 3.11.5 JIT-and-test harness.}
\label{tab:assembly_ablation_raw}
\small
\setlength{\tabcolsep}{4pt}
\begin{tabular}{lrrrrrrr}
\toprule
 & \multicolumn{4}{c}{Functional outcome} & \multicolumn{3}{c}{Compilation outcome} \\
\cmidrule(lr){2-5}\cmidrule(lr){6-8}
Condition & \passat{1} & \passat{5} & \passat{10} & Any-pass & \compileat{1} & \compileat{5} & \compileat{10} \\
\midrule
Matched baseline & 17.21\% & 22.21\% & 24.03\% & 37/154 & 77.34\% & 93.43\% & 94.81\% \\
Instruction shuffle & 17.92\% & 23.97\% & 26.62\% & 41/154 & 77.47\% & 94.97\% & 96.75\% \\
Cross-task substitution & 17.99\% & 23.70\% & 27.27\% & 42/154 & 78.25\% & 94.34\% & 96.10\% \\
\midrule
Archived baseline\textsuperscript{*} & 17.66\% & 24.11\% & 27.27\% & 42/154 & 76.88\% & 93.80\% & 95.45\% \\
\bottomrule
\end{tabular}
\vspace{2pt}

\begin{minipage}{0.98\textwidth}
\footnotesize\textsuperscript{*}Reproducibility reference from an earlier scoring pass on the same checkpoint; not scored in the same paired run as the ablated conditions and not used as the inferential comparator.
\end{minipage}
\end{table*}

The observed compilation rates are high and do not decrease under either intervention. Relative to the matched baseline, \compileat{1}/\compileat{5}/\compileat{10} change by $+0.13/+1.54/+1.95$ percentage points for shuffle and $+0.91/+0.91/+1.30$ for cross-task substitution. These compilation rates are descriptive; the paired tests below concern functional \passatk.

Table~\ref{tab:assembly_ablation_stats} treats both interventions and all three $k$ values as one six-test Holm family. Exact task-level sign-flip probabilities replace the earlier Monte Carlo approximation. Pointwise 95\% intervals use 100,000 paired task bootstraps (analysis seed 20260905); they are not simultaneous intervals. The exact test operates on integer numerators of the unbiased estimator, avoiding floating-point ambiguity at discrete ties.

\begin{table}[!t]
\centering
\caption{Graph ablation vs.\ matched baseline on the semantic view, seed 42. Differences are ablation minus baseline in percentage points. Exact task-level sign-flip $p$-values; Holm adjustment over all six tests; pointwise bootstrap intervals.}
\label{tab:assembly_ablation_stats}
\begin{tabular}{lcrlrr}
\toprule
Condition & $k$ & $\Delta$ (pp) & 95\% CI (pp) & Raw $p$ & Holm $p$ \\
\midrule
Instruction shuffle & 1 & $+0.71$ & $[-0.39,+1.88]$ & .2740 & .822 \\
             & 5 & $+1.77$ & $[-0.58,+4.22]$ & .1616 & .808 \\
             & 10 & $+2.60$ & $[-0.65,+6.49]$ & .2891 & .822 \\
\midrule
Cross-task substitution & 1 & $+0.78$ & $[-0.39,+2.14]$ & .2860 & .822 \\
             & 5 & $+1.49$ & $[-0.71,+3.73]$ & .1895 & .808 \\
             & 10 & $+3.25$ & $[0.00,+6.49]$ & .1250 & .750 \\
\bottomrule
\end{tabular}
\end{table}

Neither graph perturbation produces statistically detectable degradation under the semantic contract in the separate six-test Holm family. Coverage@10 has 6 gains and 2 losses for shuffle (exact McNemar $p=.2891$), and 6 gains and 1 loss for cross-task substitution ($p=.1250$). All point estimates favor the ablated condition.

In the ablation-minus-baseline direction, the pointwise \passat{10} intervals are $[-0.65,+6.49]$ points for shuffle and $[0.00,+6.49]$ for cross-task substitution. Positive upper limits describe possible improvement under ablation, not a six-point loss. The discrete bootstrap can have a zero lower endpoint while the exact two-sided test remains non-significant. These within-run intervals omit training- and decoding-seed variation; no equivalence margin was specified, and the results establish neither equivalence nor instruction non-use.

\textbf{Opaque name-only extension.} All three completed graph-ablation conditions yield zero passing candidates: 0/154 tasks and 0/4,620 candidates across the three arms (Table~\ref{tab:assembly_ablation_headline}). The observed floor permits no measurement of further decline. With no discordant outcomes, the exact randomization distribution is degenerate and its conventional two-sided $p$-value is 1, so we omit a separate significance table. The 4,620 candidates are clustered within the same 154 tasks, not independent experimental units. Zero observed passes do not establish a zero population success probability or rule out success under another seed or input contract.

For the secondary any-compile endpoint, shuffle versus baseline has 7 gains/5 losses ($p=.7744$), cross-task versus baseline 4/4 ($p=1$), and cross-task versus shuffle 4 gains/6 losses ($p=.7539$). These unadjusted exact McNemar checks are descriptive; none is significant even before adjustment.

\begin{table}[!t]
\centering
\caption{Graph ablation on the opaque name-only HE-154 view, seed 42, ten candidates/task. Observed \passatk is zero for every $k$ and condition; compilation counts vary.}
\label{tab:assembly_ablation_headline}
\begin{tabular}{lrrrr}
\toprule
Condition & \passat{10} & Any-pass & Any-compile & Compiling cands. \\
\midrule
Matched baseline (mode=\texttt{none}) & 0.00\% & 0/154 & 13/154 & 21/1,540 \\
Instruction shuffle & 0.00\% & 0/154 & 15/154 & 27/1,540 \\
Cross-task substitution & 0.00\% & 0/154 & 13/154 & 21/1,540 \\
\midrule
Archived name-only arm\textsuperscript{\dag} & 0.00\% & 0/154 & --- & 49/1,540 \\
\bottomrule
\end{tabular}
\end{table}
\noindent\textsuperscript{\dag}Construct-validity run reported in Table~\ref{tab:metadata_ablation} (\S\ref{sec:metadata_intervention}): same dataset and checkpoint SHA-256 hashes and nominal seed 42, independently generated on a different pod. \passatk is 0 in both pods; the compile counts are disclosed side by side below and are not averaged, reconciled, or used as an inferential comparator.

The new baseline compiles 21/1,540 candidates (1.36\%), compared with 49/1,540 (3.18\%) in the earlier name-only run; neither records a pass. We keep these runs separate because nominally identical seeds can produce different candidate streams across launches and environments. The cause of this compilation difference is unresolved. Every graph-ablation contrast uses its newly generated baseline within the same view and scoring environment.

The typed opaque arm retains types and arity and solves 7/154 tasks, but was not graph-ablated. Matched perturbations of that nonzero-baseline contract with repeated seeds would test residual binary dependence more directly. The present endpoint is aggregate functional success, not generated-text equality, architectural permutation invariance, or the usefulness of binary information to another policy.

These controls are separate from the exploratory 16-contrast architecture sweep in Appendix~\ref{app:architecture}. Its cross-task graph and block-order contrasts have oppositely signed point estimates and different provenance. A block-order shuffle is not the within-task instruction derangement used here, and the two studies are not treated as replications.

\subsection{Answer to RQ5}

Functional success depends strongly on the exposed metadata bundle in the paired seed-42 intervention. The signature-only comparator records performance without binary input, and the matched graph controls find no statistically detected degradation under the semantic contract and a zero floor under opaque name-only input. The practical conclusion is to report metadata-assisted reconstruction as a distinct setting and verify channel dependence before attributing success to instruction recovery. Recall versus inference and positive instruction-use attribution remain unresolved.

\section{Sampling Reliability and Task-Set Stability}
\label{sec:sampling_reliability}

RQ6 asks what changes when only the decoding seed changes. The study uses one frozen checkpoint, one MF-174 evaluation file, identical task order, identical scorer, and identical decoding parameters under seeds 42--46.

\begin{table}[!t]
\centering
\caption{Same-configuration reseeding on MF-174. Only the seed varies.}
\label{tab:reseed}
\begin{tabular}{lrrrrrrr}
\toprule
Metric & 42 & 43 & 44 & 45 & 46 & Mean & SD \\
\midrule
\passat{10} & 6 & 6 & 6 & 4 & 2 & 4.8 & 1.79 \\
\compileat{10} & 123 & 124 & 134 & 125 & 124 & 126.0 & 4.53 \\
First-candidate pass & 2 & 1 & 2 & 1 & 2 & 1.6 & 0.55 \\
\bottomrule
\end{tabular}
\end{table}

Changing only the decoding seed shifts \passat{10} by as many as four tasks across these five runs. For this model, corpus, prompt, temperature, and $k$, treatment differences of comparable size therefore need repeated, paired analysis.

The evaluator accepts all 175 MF references under compilation and functional scoring. Re-deriving candidate-level scores under a later scorer revision also reproduces the reported counts, providing a check on the archived outcomes.

\subsection{Aggregate Counts and Task Identities}

\begin{table}[!t]
\centering
\caption{Finite-sample stability on MF-174 across five decoding seeds. ``All'' is solved in every seed; ``any'' is solved in at least one. $J$ is mean pairwise Jaccard. $R_J$ compares pooled observed overlap with an independent-marginals plug-in baseline, as defined below. All columns use the same 174 tasks.}
\label{tab:task_stability}
\small
\begin{tabular}{lrrrrrr}
\toprule
 & \multicolumn{5}{c}{\passat{10}} & \multicolumn{1}{c}{\compileat{10}} \\
\cmidrule(lr){2-6}\cmidrule(lr){7-7}
Arm & Mean & All & Any & $J$ & $R_J$ & $J$ \\
\midrule
Constants removed & 2.6 & 1 & 7 & 0.342 & 46.1 & 0.844 \\
Baseline & 4.8 & 1 & 15 & 0.203 & 14.7 & 0.853 \\
Structure substituted & 5.2 & 1 & 12 & 0.302 & 20.3 & 0.829 \\
Typed contract & 10.0 & 4 & 24 & 0.375 & 12.5 & 0.973 \\
\bottomrule
\end{tabular}
\end{table}

For solved-task sets \(A_i\) with sizes \(a_i=|A_i|\), the pooled plug-in ratio is
\[
R_J =
\frac{\sum_{i<j}|A_i\cap A_j|}{\sum_{i<j}|A_i\cup A_j|}
\bigg/
\frac{\sum_{i<j}a_i a_j/N}{\sum_{i<j}(a_i+a_j-a_i a_j/N)},
\qquad N=174.
\]
Its denominator is a ratio of expected intersection and union sizes under independent fixed-size subsets, not the expected Jaccard itself. Thus \(R_J\) compares pooled overlaps, whereas the displayed \(J\) averages the ten pairwise Jaccards.

Stable aggregate counts conceal substantial solved-task turnover. The typed-contract arm has the tightest count ($10.0\pm1.0$), yet its five runs collectively solve 24 tasks and solve only 4 in every run. The baseline solves 15 tasks at least once but only 1 in every seed. These all/any counts identify small stable task cores at the observed success rates.

The compiling sets from the same generations are more stable in absolute terms. Together with the archived-score replay, this pattern is consistent with sampling-sensitive functional success. It does not rule out every runtime, scorer, or evaluation-order effect outside the replayed candidates.

\subsection{Implications for Task-Level Analyses}

Aggregate \passatk supports configuration-level comparisons through paired repeated-seed designs. Task-level explanations need a further check: whether the same functions recur in the solved set. Accordingly, the historical length and error analyses in Section~\ref{sec:error} describe one draw; the reseeding study supplies the method for assessing persistent task-level changes.

Reporting the numbers solved in every seed and in any seed, or the full solve-frequency distribution, makes this stability visible. These are finite-sample indicators rather than estimates of latent capability. The turnover pattern is consistent with the heavy-tailed account of inference scaling: per-task failure rates fall exponentially with the number of attempts, while a heavy-tailed distribution of single-attempt success probabilities produces an aggregate power law~\citep{schaeffer2025monkeys}, with related models of coverage growth and budget allocation~\citep{levi2024inference,kazdan2025efficient,meir2026red}. Tasks with low success probability enter and leave the solved set across seeds while the aggregate count moves comparatively little.

\subsection{Answer to RQ6}

Under the studied MF protocol, aggregate counts are substantially more stable than the identities of solved tasks. Repeated-seed paired inference is required for small treatment effects, and repeated task identities are required before attributing improvements to particular function classes.

\section{Input-Channel Interventions}
\label{sec:channel_integrated}

RQ7 examines the contributions of typed contracts, constants, and structural input to functional correctness and interface conformance on MF-174. This separate system and corpus complements the HumanEval metadata intervention rather than replicating it.

\subsection{Corpus Dependence}

Corpus stratification locates the typed-contract compilation gain in the generated tasks. MF-175 contains 136 model-generated and 39 legacy tasks; the one contaminated task is removed from replicated analyses. Over the five blocks retaining task-level provenance, the aggregate effect is $+41.4$ tasks: $+42.2$ from the 136 generated tasks and $-0.8$ from the 38 clean legacy tasks. The correctness-by-origin interaction was not measured, so this stratum-level conclusion applies to compilation only.

\subsection{Channel Results}

\begin{table}[!t]
\centering
\caption{Input-channel interventions on MF-174. Baseline, typed contract, and structure substitution have six observed seed blocks; constants removal has five; the no-binary oracle has two. Only the prospectively extended typed-contract \passat{10} contrast is treated as supportive inferential evidence.}
\label{tab:channel_integrated}
\small
\begin{tabular}{lrr}
\toprule
Input & \passat{10} & \compileat{10} \\
\midrule
Baseline & $5.0\pm1.67$ & $125.8\pm4.07$ \\
Typed contract & $\mathbf{10.3\pm1.21}$ & $\mathbf{168.0\pm2.19}$ \\
Constants removed & $2.6\pm1.14$ & $124.2\pm4.66$ \\
Structure substituted & $5.0\pm1.79$ & $117.2\pm2.93$ \\
Typed contract, no binary & $4.0$ (5, 3) & $173.5$ (174, 173) \\
\midrule
\multicolumn{3}{l}{Effect versus baseline; exact paired sign flip where applicable} \\
Typed contract & $+5.3,\ p=0.031$ & $+42.2,\ p=0.031$ \\
Constants removed & $-2.2,\ p=0.25$ & $-1.8,\ p=0.50$ \\
Structure substituted & $0.0,\ p=1.00$ & $-8.7,\ p=0.031$ \\
\bottomrule
\end{tabular}
\end{table}

\subsubsection{Typed contract.}
Replacing the opaque signature with \fnz and declared parameter and return types raises \compileat{10} from $125.8$ to $168.0$ and \passat{10} from $5.0$ to $10.3$. On seed 42, candidate-level compilation rises from 578/1,740 to 1,323/1,740, while diagnostics attributable to wrong arity or types fall from 1,081 to 312.

All six observed \passat{10} block differences are positive: $(3,5,4,5,9,6)$. The exact sign-flip result is $2/2^6=0.031$. A six-block bootstrap estimates a mean difference of $+5.3$ tasks with 95\% CI $[+4.0,+7.0]$; a task-by-seed hierarchical bootstrap over the five blocks retaining task-level records gives $+5.8$ $[+1.4,+10.6]$. This is supportive evidence under the sixth-seed extension chronology described below, not an independent confirmation.

In the relevant single-seed comparison, \passat{10} reaches 14 with typed-contract training, 10 with the contract supplied only at inference, and 6 at baseline. The ordered point estimates motivate testing both information supply and training adaptation; the available paired test does not resolve their individual contributions.

\subsubsection{Constants removed.}
Deleting recovered strings and numeric literals while leaving the structural payload unchanged moves \passat{10} by $-2.2$ tasks and \compileat{10} by $-1.8$ over five seeds. Neither change is resolved by the paired tests. On seed 42, the 96 tasks with literals fall from four solved to zero; the 79 no-literal tasks produce byte-identical predictions, verifying that this intervention leaves no-literal inputs unchanged.

\subsubsection{Structure substituted.}
Length-matched cross-task substitution lowers \compileat{10} in every observed block, while functional directions are mixed. Each MF task receives another task's F2 structural body (assembly and CFG) and retains its target-side recovered prefix and constants. Because both semantics and structural features change, this tests the combined structural payload rather than instruction use alone; it is distinct from the HumanEval within-task graph shuffle.

\subsection{Prospective Sixth-Seed Extension}

The five-block design has a minimum two-sided exact $p$ above $0.05$. After inspecting those blocks, the authors sealed one additional seed, one endpoint (typed contract versus baseline on \passat{10}), and a hard stop. The amendment also recorded a uniformly reduced generation batch and a correction to a chunked-sampling defect before seed-47 scoring. Seed 47 returned a $+6$-task difference. The six-block analysis is supportive: the extension followed inspection of the initial pattern, and the combined analysis includes a diagnostic seed.

\subsection{Withholding the Binary}

The no-binary oracle keeps the typed contract byte-for-byte and replaces F2 with the empty string. It was prospectively specified for two seeds. The arm reaches 174 and 173 compiling tasks of 174, and 5 and 3 functionally solved tasks.

\begin{table}[!t]
\centering
\caption{Candidate-level rates on MF-174, seed 42.}
\label{tab:no_binary_candidates}
\begin{tabular}{lrr}
\toprule
Input & Compiled & Passed \\
\midrule
Baseline & 33.2\% & 0.52\% \\
Typed contract & 76.0\% & 1.15\% \\
Typed contract, no binary & 89.0\% & 0.57\% \\
\bottomrule
\end{tabular}
\end{table}

The candidates are diverse (9.95--9.97 distinct outputs per task) and fail primarily at runtime assertions rather than static typing. This directly separates near-perfect interface conformance from correct computation. The arm remains a gold-derived oracle: it assumes perfect type and arity recovery and supplies an out-of-distribution empty binary payload.

\subsection{Answer to RQ7}

Typed contracts strongly affect interface conformance and yield a supportive repeated-seed functional gain on MF-174. Cross-task structural substitution reduces compilation with no detected correctness effect, and constants removal remains unresolved. The no-binary oracle shows that interface conformance can saturate while functional success stays low. The channel results therefore identify different requirements for compiling and passing tests in this synthetic-majority corpus.

\section{Evaluation Provenance and Post-Mortem}
\label{sec:provenance}

A post-submission audit found that a historical prompt builder extracted up to ten assertion lines from each HumanEval-Dart row's \texttt{tests} field and inserted them into the policy prompt. The evaluator later appended the same tests to generated code. All 154 rows were affected, 38 saturated the ten-line cap, and 133/154 exposed their entire test suite within the prompt window. Truncation preserved the fixed prefix and removed assembly first, so the leaked scoring evidence survived precisely where instruction evidence did not.

We exclude every affected historical functional pool from the findings. The exclusion is deliberately conservative: the runs lack pinned prompt hashes, source commits, and dataset hashes, so file timestamps cannot distinguish clean pools and some potentially unaffected pools are also excluded. The prior fine-tuning corpora used for the original TOSEM results do not contain this specific scoring-test field. That observation identifies the boundary of the discovered defect, not a general absence-of-contamination guarantee.

The metadata-intervention results postdate the correction and use the test-hint-free contracts described in Section~\ref{sec:metadata_intervention}; the audit field \nolinkurl{rows_leaking_scoring_tests} is zero. The replication package records per-run prompt schemas, corpus and checkpoint identities, candidate-level scores, contamination notices, and replay reports.

\begin{table*}[!t]
\centering
\caption{Result-lineage summary. Detailed machine-readable records are in the artifact.}
\label{tab:lineage}
\scriptsize
\setlength{\tabcolsep}{2pt}
\begin{tabularx}{\textwidth}{>{\raggedright\arraybackslash}p{0.18\textwidth}>{\raggedright\arraybackslash}p{0.18\textwidth}>{\raggedright\arraybackslash}p{0.15\textwidth}>{\raggedright\arraybackslash}p{0.13\textwidth}Y}
\toprule
Evidence & Prompt/data lineage & Candidate replay & Status & Use in article \\
\midrule
Original fine-tuning study & CB-73, CP-126, HE-154; task 160 forced zero & per-task summaries, not complete candidates & retained with limits & Configuration-specific RQ1--RQ3; historical semantic-signature condition \\
Aligned metric audit & HE-154/HE-141; same candidates for all metrics & candidate replay for seeds 42--44 & retained with global-audit caveat & Secondary RQ4 association analysis \\
Metadata intervention & clean prompt schema; HE hashes and overlap audit & candidate-level scoring retained & retained & Primary construct-validity evidence \\
MF reseeding and channels & sealed F2 grammar; one contaminated task excluded & task-level records for seeds 42--46; aggregate block for 47 & retained with limits & Replicated secondary and exploratory evidence \\
Historical test-leaking pools & prompt/test provenance incomplete & not trusted for inclusion & withdrawn & Audit record only \\
Architecture sweep & planned seeds incomplete; provenance header reports FAIL; aggregate/replay disagreements & candidate replay treated as canonical & exploratory; withdrawn from claims & Appendix~\ref{app:architecture} only \\
Initial frontier 5/60 run & truncation, closure omission, non-content-addressed cache, unknown effective $k$ & invalidated & withdrawn & Appendix~\ref{app:frontier} lineage only \\
Replacement frontier run & paired manifest, complete 1,750 slots, hash-pinned artifacts & complete & exploratory reference & Appendix~\ref{app:frontier} \\
\bottomrule
\end{tabularx}
\end{table*}

The post-mortem provides a concrete provenance requirement: bind each score to its prompt contract, test boundary, model, task order, denominator, and candidate-level evidence. This makes contamination checks and archived-outcome replay part of the evaluation protocol rather than a separate documentation exercise.

\section{Discussion}
\label{sec:discussion}

\subsection{Unified Interpretation}

The studies connect four aspects of decompilation evaluation. Corrected functional inference distinguishes the two Qwen3-8B regressions from four inconclusive adapter outcomes. Common-candidate scoring quantifies modest association between static similarity and correctness and weak discrimination by compilation coverage. The metadata intervention demonstrates a large within-run dependence on the supplied specification. Repeated decoding then shows why aggregate coverage and solved-task identity need separate reliability measures.

Together, these findings support an evaluation protocol that identifies (i) available metadata, (ii) supplied binary information, (iii) the behavior checked by each metric, (iv) the sampling design, and (v) the evidence available for replay. This protocol makes model comparisons more informative by specifying what a measured gain or regression represents.

\subsection{\passatk Is Necessary but Not Sufficient}

Execution distinguishes functional outcomes among compiling candidates. In the aligned audit, \codebleu has modest association with correctness and compilation is nearly saturated. The meaning of \passatk is specific: success against the available harness under the stated prompt and sampling contract.

For observed any-success coverage, the operational statement is:
\begin{quote}
Under the stated metadata, binary, test, and sampling contract, at least one of $k$ generated candidates satisfied the available harness.
\end{quote}
The unbiased \passatk estimator summarizes success probability over candidate subsets. Matched metadata and binary-input controls address instruction-use attribution; repeated task identities support function-class analyses; and test adequacy determines the behavioral scope.

\subsection{Metadata-Assisted Reconstruction Is a Distinct Setting}

Symbols, decompiler-inferred types, selector information, and partial debug metadata are useful inputs in real reverse-engineering settings. Metadata-assisted reconstruction measures performance when those inputs are available; metadata-scrubbed reconstruction measures performance after their removal. Reporting both conditions makes that engineering distinction explicit. Within the scrubbed condition, matched binary-input controls are still needed to separate instruction use from task priors and interface information.

\subsection{Configuration-Specific Effects of Task-Specific Fine-Tuning}

The historical matrix shows why adaptation should be evaluated against its own base: two Qwen3-8B adaptations regress at every $k$, while the four other variants are inconclusive after Holm correction. These are configuration-specific outcomes. The bases differ in architecture, pretraining, reasoning/distillation, and baseline performance, while their exact historical training paths remain unresolved; language and compiler optimization also vary across mixtures. The independent graph-prefix training at seeds 42--44 supplies separate evidence about metric association and training-seed variation, not a replication of v1--v6 or an explanation based on scale.

Changes in reasoning behavior, decoding diversity, or particular task skills are possible mechanisms for future testing. Assessing them would require a verified reasoning-trace serialization path for the six runs and repeated task-level measurements. The current evidence identifies functional outcomes without selecting among those mechanisms.

For practice, establish an unadapted baseline, score the same candidates with static, compilation, and execution measures, repeat training and decoding where feasible, and use multiplicity-corrected functional evidence to assess improvement. The Qwen3-8B regressions demonstrate the value of this check for the evaluated adaptations.

\subsection{Implications of the Input-Channel Study}

The channel study explains an important distinction between compilation and correctness. Supplying the typed interface checked by the front end raises compilation sharply. With the binary removed and the typed contract retained, compilation nearly saturates while functional success stays low. Under this corpus and prompt, supplying the interface resolves many front-end failures without resolving the computation.

Stratification further shows that the compilation gain comes from the generated-task stratum. The structure intervention changes several features together, and the constants effect remains unresolved. The methodological contribution is to measure channels and corpus strata separately so that an aggregate result can be traced to the conditions producing it.

\subsection{Why the Complete Research Record Matters}

The secondary investigations provide useful design and feasibility information. Rejection-sampled fine-tuning records a diversity-collapse mode; out-of-memory RL and unavailable distillation APIs identify infrastructure requirements; the architecture audit documents the basis for excluding its conclusions; and the frontier and ARM64 runs provide preliminary reference points. Keeping these records with explicit status distinguishes observed outcomes from unmeasured efficacy and helps future work avoid repeating known problems.

\section{Threats to Validity}
\label{sec:threats}

\subsection{Internal Validity}

\textbf{Optimization mismatch.} Dart training binaries are AOT optimized while Swift is compiled at \texttt{-O0}. The observed 4B Dart--Swift contrast combines possible language competition, optimization-level distribution shift, and corpus-composition effects. Their individual contributions are not identified.

\textbf{Hyperparameter sensitivity.} LoRA/DoRA rank, learning rate, dropout, data size, and training duration were not exhaustively optimized per model. Different settings could reproduce, reduce, or reverse an observed effect; the historical findings concern the evaluated configurations.

\textbf{Historical single outcomes and follow-up retraining.} Each original v1--v6 variant has one archived training outcome. The only repeat within that historical matrix is an inference-only Qwen3-8B/v5 pair without complete seed or candidate provenance, and public candidate scripts conflict with the submitted hyperparameters. By contrast, the graph-prefix follow-up records independent training seeds 42--44 and distinct checkpoint hashes for matched configurations. Conclusions about the historical adapters therefore remain configuration-specific, while seed variability can be estimated for the controlled follow-up.

\textbf{Legacy forced-zero rule.} Task 160 was assigned $c=0$ without execution. Its missing outcome can change an absolute score by up to $+0.65$ points and a paired difference by up to $\pm0.65$ points. The exclusion and worst-case checks in Appendix~\ref{app:benchmark_audit} preserve the v5/v6 regression decisions. They bound the missing outcome rather than reconstructing its execution.

\textbf{Single-seed metadata intervention.} The 42-to-7 metadata result is a large paired within-run contrast, not a seed-replicated estimate. Its magnitude across seeds and the identities of retained tasks are unknown.

\textbf{Off-distribution opaque contracts.} The typed-opaque and opaque name-only prompts differ lexically from the training contract. Higher typed-opaque compilation argues against a general inability to generate Dart. The name-only arm additionally removes types and arity and has low compilation. Semantic removal and its lexical or distributional consequences are not fully separated by this design.

\textbf{Name-only and signature-only controls.} The opaque \fnz arm retains the binary; the no-binary comparator receives the full semantic signature and differs in adaptation. Neither isolates the semantic name alone. Graph perturbations cover the semantic and opaque name-only views, with a zero floor in the latter; the typed opaque arm with seven covered tasks was not graph-ablated. Instruction-use attribution therefore remains unresolved. Section~\ref{sec:assembly_ablation} also distinguishes verified input mappings from incomplete loaded-state attestation.

\textbf{Retrospective extension.} The MF typed-contract study added seed 47 after the first five blocks were inspected. The endpoint and hard stop were sealed before that seed was generated, but the six-block test is supportive rather than an independent prospective replication. Seed 47 does not retain the same task-level provenance as seeds 42--46 for the hierarchical bootstrap.

\textbf{Historical task labels.} The original error and length analyses use task identities from one historical generation set. Repeated-seed turnover on MF-174 shows why those identities should be interpreted descriptively.

\textbf{Training-seed audit provenance.} The selected no-edge graph-prefix runs at seeds 42--44 have distinct checkpoint identities and use the expected encoder revision. The enclosing 38-run report remains \texttt{FAIL} for two CLAP encoder-revision mismatches; aggregate/replay disagreements are a separate consistency issue. Table~\ref{tab:aligned_metrics} uses canonical archived candidate labels, whereas Table~\ref{tab:architecture_core} applies the documented stability overlay.

\subsection{Construct Validity}

\textbf{Metadata bundle.} The principal intervention removes function names, parameter names, helper and disassembly symbols, and a source path. It does not identify the effect of any single field.

\textbf{Recall versus inference.} A descriptive identifier may retrieve a memorized solution or serve as a specification for fresh inference. The measured metadata effect includes both possibilities. Paraphrased names, misleading names, and post-cutoff tasks would help distinguish them. The five-criterion training-overlap check covers the identified fine-tuning corpus, not all semantic duplicates or pretraining exposure.

\textbf{Test adequacy.} \passatk measures satisfaction of the available harness, not complete semantic equivalence. HumanEval-Dart has a median of seven assertions, 12 reference failures, and four contract defects. The audited denominators correct certification status but do not prove that every passing generated candidate is correct on untested inputs.

\textbf{\compileatk semantics.} Compilation under this harness includes interface conformance and type consistency. It should not be described as generic syntactic correctness or used as a proxy for behavioral correctness.

\textbf{\codebleu implementation.} The Dart implementation uses AST node-type overlap and an identifier-level data-flow heuristic. These static components can remain high despite behavior-changing token errors.

\subsection{Statistical Conclusion Validity}

The historical \codebleu, CP-126, and HE-154 panels contain 73, 126, and 154 tasks, respectively, with different candidate pools. Same-candidate metric association and discrimination are evaluated separately on HE-141 for one configuration and three training seeds. The historical \codebleu inferential family was not preregistered and is treated as exploratory. For the principal variant--base functional family, Holm correction spans 18 tests; v5 and v6 regressions remain significant, while the other four comparisons are inconclusive. Five-block sign-flip tests have a minimum two-sided $p$ of 0.0625. Non-significance establishes neither absence of an effect nor equivalence. Secondary channel contrasts are exploratory, and the graph-ablation intervals omit between-seed variation.

\subsection{External Validity}

The article studies one target language, x86-64 in the principal experiments, and a limited set of model families. Although the compiler mode is used for production deployment, HumanEval-Dart contains short algorithmic functions rather than production application logic. MF-174 is majority model-generated, and its large compilation effect is concentrated in that stratum. The legacy MF stratum lacks original repository and licence provenance. The ARM64 Flutter run retains semantic metadata, uses model-generated sources, and has one seed, so it is not a production-code or conceptual replication. The frontier measurement represents one provider, model, date, and decoding configuration.

\subsection{Provenance and Researcher Degrees of Freedom}

The research programme spans multiple checkpoints, corpora, toolchain revisions, and historical prompt builders, creating retrospective selection risk. The evidence taxonomy and claim ledger make those choices visible. The architecture sweep has incomplete planned seeds and a failed provenance audit and is retained as exploratory history; historical test-leaking pools are excluded from findings. The replacement artifact pins prompt schemas, task order, hashes, candidate-level scores, and exclusions. Unresolved legacy licensing limits raw-material redistribution, and missing historical records limit end-to-end reproduction.

\subsection{Generative AI Use}

Generative systems assisted with prose, scripts, audit organization, and adversarial review. ChatGPT-5.5 with \texttt{xhigh} reasoning was also used during the original HumanEval-Dart construction to translate HumanEval implementations and tests into Dart and to execute the translated tasks for validation. The original conversation and run transcript were not retained; the present benchmark certification is independent and reports the defects it finds. These uses create risks of translation, citation, and interpretation error. The authors report the systems and affected activities, regenerate statistics from artifacts, verify citations against primary sources, and take responsibility for every claim and benchmark row.

\section{Evaluation Recommendations}
\label{sec:recommendations}

The combined evidence supports the following reporting protocol.

\begin{enumerate}[leftmargin=*,itemsep=2pt,topsep=2pt]
\item \textbf{Report metadata-assisted and metadata-scrubbed conditions.} Remove function and parameter names, helper and disassembly symbols, and source-path cues. When metadata is legitimately present, report it as a separate condition rather than conflating it with stripped-binary recovery.
\item \textbf{Include a signature-only baseline.} Measure performance when the model receives the exposed specification without the binary. An unadapted base provides a low-cost reference; isolating the incremental binary contribution requires matching adaptation and all other prompt fields.
\item \textbf{Define and align each metric operationally.} State whether compilation checks interface conformance, whether surface metrics are conditioned on compiling candidates, and what tests \passatk executes. Evaluate associations on the same tasks and candidate outputs; separate-panel aggregate movements do not estimate same-candidate associations.
\item \textbf{Certify the benchmark.} Execute references against their own harnesses, enumerate contract defects, report clean denominators, and audit training/evaluation overlap.
\item \textbf{Replicate training and stochastic decoding.} Distinguish independent training seeds from repeated inference over one checkpoint. Report paired per-seed effects, intervals, the exact test implied by the blocking structure, and a declared multiplicity family. Do not interpret a single draw as configuration-level robustness.
\item \textbf{Report task-set stability.} Give all-seed, any-seed, or full solve-frequency counts before claiming that a method fixes particular tasks, failure modes, or complexity strata. A beta-binomial fit over per-task solve frequencies summarizes this distribution compactly~\citep{kazdan2025efficient}.
\item \textbf{Stratify by provenance.} Separate model-generated benchmarks, human-authored benchmarks, legacy code, and production binaries; aggregate effects may be driven by one stratum.
\item \textbf{Fail closed on provenance.} Pin prompt schemas, model and corpus hashes, toolchain revisions, task order, candidate-level scores, and denominators. Refuse partial or budget-truncated scores.
\item \textbf{Include information-removal controls.} Specify whether ``name only'' means a neutral contract with binary input or a semantic name without binary input. Match adaptation and prompt contracts, verify that perturbations reach the model's consumed input, and retain a nonzero baseline where possible. Report zero-floor outcomes as floors, not equivalence or channel irrelevance. Typed-opaque graph perturbations with repeated seeds are the next direct test for the present system.
\item \textbf{Classify unsuccessful programmes correctly.} Distinguish statistically unresolved contrasts, discrete-test limits, feasibility failures, invalidated runs, and withdrawn evidence.
\end{enumerate}

\section{Future Work}
\label{sec:future}

The results identify the following priorities for further evaluation.

\textbf{Mechanism-discriminating metadata controls.} Paraphrased, misleading, and post-cutoff identifiers could separate exact benchmark recall from inference over descriptive metadata.

\textbf{Optimization-controlled cross-lingual training.} A factorial comparison that varies language and optimization regime, with matched data and token budgets, would help separate their effects. Merely enabling Swift optimization would not make its compiler transformations identical to Dart's.

\textbf{Task-frequency rather than binary solvability models.} Repeated generations could replace one-run solved/unsolved labels with estimated solve frequencies in complexity and error analyses.

\textbf{Hierarchical decompilation.} Chunking long assembly into basic blocks, loops, or regions may address the observed length degradation.

\textbf{Production and cross-ISA replication.} The FL-343 infrastructure separates signature text from binary soft-prefix embeddings. Applying metadata-scrubbed contracts and repeated seeds to non-synthetic application code would extend the current preliminary observation to the intended deployment setting.

\textbf{Execution-guided optimization.} Rejection sampling in the present programme did not improve \passat{10}, while RL could not be executed under the available memory budget. Future systems could test execution feedback with infrastructure sized for the intended objective rather than treating feasibility failures as evidence about efficacy.

\section{Conclusion}
\label{sec:conclusion}

This study contributes an execution benchmark, corrected functional comparisons, and an auditable evaluation protocol for Dart AOT decompilation. Paired inference across \passatk at $k=1,5,10$ identifies regressions in both Qwen3-8B adapters after Holm correction; the other four historical comparisons are inconclusive.

The separate three-seed common-candidate audit supplies direct association and discrimination evidence beyond the historical matrix's separate panels. On HE-141, best \codebleu has modest association with \passat{10} ($\rho=.218$--$.246$), \compileat{10} has weak association ($\rho=.072$--$.082$), and 21.0--23.3\% of compiling candidates pass. Execution distinguishes functional success among candidates that compile.

The metadata intervention establishes a large within-run effect of the input contract. Removing semantic names and related cues as a bundle while preserving types, arity, and instruction content reduces coverage@10 from 42 to 7 tasks in the paired seed-42 run, with 35 losses and no gains. The distinct opaque name-only/binary arm solves 0/154 tasks, and the unadapted semantic-signature-only/no-binary comparator solves 36/154. The latter is an adaptation-unmatched reference, not an isolated binary-effect estimate.

Matched graph perturbations show no statistically detectable degradation under the semantic contract and a zero floor under opaque name-only input. Instruction-use attribution remains unresolved; repeated perturbations of the nonzero typed-opaque condition are the next test. These controls complement the measured metadata sensitivity by identifying the comparison needed to assess residual binary dependence.

Across five decoding seeds on MF-174, the baseline solves 4.8 tasks on average, 15 at least once, and one in every seed. This solved-task turnover is distinct from the follow-up training-seed variation measured at seeds 42--44. Task-160 sensitivity checks preserve the historical Qwen3-8B regressions. The artifact supplies six graph-ablation prediction/score pairs, hash-matched source snapshots, and exact statistical replay, with historical provenance and release limits documented explicitly. The resulting protocol is practical: certify benchmarks, align metrics, separate metadata and binary channels, repeat training and inference, correct multiplicity, and report both task-set stability and replay scope.

\appendix

\section{Complete Prompt-Channel Audit}
\label{app:prompt_audit}

The literature audit searched published neural-decompilation systems and benchmarks from 2024--2026 for which at least one of the following was public: prompt-construction code, a released evaluation dataset, or a verbatim prompt template. Configurations whose prompts could not be established were recorded as unverifiable and excluded rather than inferred. Each included row was checked against its released artifact independently of the search that located it. The systems are LLM4Decompile~\citep{tan2024llm4decompile}, SALT4Decompile~\citep{wang2025salt4decompile}, SK2Decompile~\citep{tan2025sk2decompile}, and SCDBench~\citep{qin2026scdbench}.

\begin{table*}[!t]
\centering
\caption{Full audit of the six verifiable configurations. ``Original'' means semantic source metadata reaches the model; ``inferred'' means a classical decompiler supplies a declaration.}
\label{tab:prompt_audit_full}
\scriptsize
\setlength{\tabcolsep}{2pt}
\begin{tabularx}{\textwidth}{>{\raggedright\arraybackslash}p{0.22\textwidth}>{\raggedright\arraybackslash}p{0.10\textwidth}>{\raggedright\arraybackslash}p{0.11\textwidth}Y>{\raggedright\arraybackslash}p{0.10\textwidth}c}
\toprule
System / split & Function name & Parameter names & Signature or symbol channel & Metadata control & Seeds \\
\midrule
LLM4Decompile end-to-end / HumanEval-Decompile & \texttt{func0} & none & no source signature & no & 1 \\
LLM4Decompile refinement / HumanEval-Decompile & \texttt{func0} & \texttt{param\_N} & Ghidra-inferred declaration & no & 1 \\
LLM4Decompile / GitHub2025 and ExeBench & original symbol & none & demangled C++ types/arity may precede instructions & no & 1 \\
SALT4Decompile / Decompile-Eval and MBPP & generic \texttt{func} & \texttt{var1..} & no source signature & no & 1 \\
SK2Decompile / stripped split & \texttt{sub\_XXXX} & \texttt{a1,a2} & IDA-inferred declaration & no & 1 \\
SCDBench / EVM-Solidity & original & none & original types and arity; EVM selectors are part of bytecode & no & 1 \\
\bottomrule
\end{tabularx}
\end{table*}

The audit identifies two patterns in the six verifiable configurations. Identifier exposure varies by benchmark split and input channel: HumanEval-Decompile neutralizes symbols, while real-world splits can preserve true names in assembly consumed by the same system. None of the audited configurations reports a metadata-scrubbed arm or signature-only baseline. The sampling frame is the auditable configurations, not all languages or unpublished systems.

\section{HumanEval-Dart Audit Details}
\label{app:benchmark_audit}

The retained JSONL contains 154 unique HumanEval IDs. The retrospective author account describes asking ChatGPT-5.5 with \texttt{xhigh} reasoning to translate and execute the tasks for correctness checks. The construction conversation, execution transcript, and reject log were not preserved, so per-task validation and rejection decisions remain unverified. The certification below independently tests the retained ports.

Relative to IDs 0--163, the absent IDs are 23 (\texttt{strlen}), 25 (\texttt{factorize}), 32 (\texttt{find\_zero}), 38 (\texttt{decode\_cyclic}), 50 (\texttt{decode\_shift}), 103 (\texttt{rounded\_avg}), 125 (\texttt{split\_words}), 137 (\texttt{compare\_one}), 148 (\texttt{bf}), and 162 (\texttt{string\_to\_md5}). Seven---32, 38, 50, 103, 125, 137, and 148---are also absent from MultiPL-E's 157-task Dart port. Their HumanEval signatures have missing or unsupported type annotations under that translator: task 32 has a lowercase \texttt{list} annotation and no return annotation; tasks 38 and 50 lack return annotations; and tasks 103, 125, 137, and 148 have untyped parameters. HE-154 is exactly the MultiPL-E Dart ID set minus 23, 25, and 162. These observations document membership and an upstream compatibility explanation; they establish neither code lineage nor the local reason for any of the ten exclusions. An LLM-assisted port need not follow an automated translator's rejection rules.

The historical benchmark includes 154 Dart ports with test harnesses. The four contract-defective tasks are:
\begin{itemize}
\item \#121 and \#153: tests bind an undefined wrapper;
\item \#127: tests pass record arguments to a \texttt{List<int>} contract;
\item \#161: unintended string interpolation in the test source.
\end{itemize}
Twelve supplied references fail their own harness; the complete task IDs and diagnostics are retained in the machine-readable manifest. Three tasks belong to both defect populations.

The artifact records the exact compiler, test hashes, reference outcomes, normalized hidden-test digests, and membership of HE-154, HE-150, HE-142, and HE-141. The metadata intervention's 35-loss/0-gain discordance is identical on every denominator. The five training-overlap checks are exact source, identifier-normalized source, exact assembly, function-name match, and 7-gram Jaccard at a threshold of 0.8; all return zero detected overlap against the 2,496 rows behind the checkpoint.

\paragraph{Task-160 sensitivity.} The original scorer assigned \texttt{doAlgebra} $n=10,c=0$ without generation or execution. Removing this common zero gives $N=153$: every paired mean difference is multiplied by $154/153$, and exact paired sign-flip $p$-values and their 18-test Holm decisions are unchanged. This is a score-subset analysis, not a corrected 154-task experiment.

We also consider every possible missing count $c\in\{0,\ldots,10\}$ for both arms at each $k$. The maximum change in a paired mean is $\pm100/154$ points. For each contrast, the artifact calculates the largest exact two-sided $p$-value over these possible completions and applies the conservative factor 18. The worst-case family-wise bounds remain below .05 for v5 and v6 at all three $k$ values (largest bounds .0043 and .0071, respectively). The two regression findings therefore survive every task-160 completion under this score model. This sensitivity analysis addresses the forced zero; it neither recovers missing candidates nor constitutes a full benchmark repair.

\section{Full Reseeding and Channel Chronology}
\label{app:channel_chronology}

Seed 42 predates the replicated channel programme and served as diagnostic context. Seeds 43--46 are fresh blocks. Excluding seed 42 changes no reported direction. The typed-contract-minus-baseline \passat{10} differences across seeds 42--47 are $(3,5,4,5,9,6)$. The sixth block was generated only after sealing the endpoint, seed, operational deviations, tokenizer identity, and hard stop.

The constants intervention was not extended because its five-block vector contains zero-difference blocks that a sixth sign could not convert into a significant two-sided sign test. The structure-compilation contrast is sign-consistent across six blocks but was not the registered primary endpoint. All secondary channel $p$-values are reported descriptively and without a family-wise multiplicity claim.

The no-binary oracle registered expected ranges of 160--172 compiling tasks and 1--3 solved tasks on the original 175-task denominator. Observed leakage-clean values were 174/173 compiling and 5/3 solved. The registered branch classified this as substantial type- or benchmark-prior performance. The arm is retained because it demonstrates that interface conformance can saturate under a fixed contract even when functional correctness remains near baseline scale.

\section{Cross-System Reference Points and Objective-Side Investigations}
\label{app:secondary}

\subsection{Cross-System Reference Points}

\begin{table}[!t]
\centering
\caption{Descriptive reference points on the full 175-task name-scrubbed corpus, predating the one-task exclusion. Model, scale, training, and input all vary; the table supports no causal comparison.}
\label{tab:cross_system}
\begin{tabular}{llrr}
\toprule
System & Input & \passat{10} & \compileat{10} \\
\midrule
Decoder-only 8B & assembly only & 1 & 103 \\
Decoder-only 8B & recovered constants & 4 & 107 \\
Frontier reasoning model & recovered constants & 4 & 86 \\
Decoder-only 8B & oracle constants & 10 & 121 \\
Encoder--decoder & assembly + CFG & 7 & 124 \\
Encoder--decoder & typed contract & 14 & 171 \\
\bottomrule
\end{tabular}
\end{table}

The reference points illustrate performance under different information channels and model configurations. Equal aggregate counts at these low rates need not imply equal behavior; the frontier and decoder solved sets were not compared.

\subsection{Objective-Side Investigations}

Expanded synthetic supervised fine-tuning changes \passat{10} by 0.0065; its interval is $[-0.020, 0.039]$. Rejection-sampled fine-tuning spans six checkpoints, three arrangements (single pass, sequential stacking, and a folded pass over the harvested union), and two replay conditions. Across those runs, \passat{10} ranges from 12 to 18 against a baseline of 14. Sequentially stacking a second pass reduces \passat{10} from 18 to 12 while leaving \compileat{10} and first-candidate success unchanged or higher. Mean distinct candidates fall from 10.00 to 9.45. Folding the same data into one pass removes the diversity collapse (9.98 distinct candidates) but produces no gain.

Seven reinforcement-learning launches encounter hard out-of-memory errors at approximately 94\,GiB before completing a training step. The available top-$k$ API also lacks the information required for dense distillation. These attempts identify infrastructure limits; neither produces an efficacy measurement.

An oracle that extracts constants by regular expression over gold source raises \passat{10} from 0.57\% to 5.71\% with \compileat{10} at 69.1\%, demonstrating the value of supplied constants in that comparison. Gold-source access makes it an oracle rather than a deployable extractor. The binary-grounded extractor reaches 2.29\%, completes 1,746 of 1,755 extraction runs over the full function pool, and obtains micro string recall 0.465 against its own 0.8 acceptance threshold.

\section{Exploratory Architecture Investigations}
\label{app:architecture}

The architecture investigations compare graph-prefix variants and component interventions and document their provenance. The selected replay-consistent core rows support the secondary metric and seed-variation audit. Architectural conclusions are excluded from the main findings because planned replication is incomplete and the enclosing provenance audit reports failure.

\subsection{Three-Seed Core Ablation}

\begin{table}[!t]
\centering
\caption{Three-seed graph-prefix core ablation, mean $\pm$ sample SD over seeds 42--44 on 154 tasks, after the documented pass-stability overlay. The no-edge covered-task counts are 39, 37, and 44, not the unoverlaid counts 39, 37, and 45; see Section~\ref{sec:metric_finding}.}
\label{tab:architecture_core}
\begin{tabular}{lrrr}
\toprule
Variant & pass@1 & pass@10 & compile@1 \\
\midrule
Prefix, no edges & $.160\pm.010$ & $.260\pm.023$ & $.756\pm.007$ \\
Prefix + CFG & $.159\pm.014$ & $.255\pm.014$ & $.747\pm.026$ \\
Prefix + CFG + DFG & $.154\pm.018$ & $.255\pm.027$ & $.732\pm.086$ \\
\bottomrule
\end{tabular}
\end{table}

Hierarchical bootstrap intervals are:
\begin{itemize}
\item CFG minus no edges: $-0.0043$ $[-0.046,+0.033]$;
\item CFG+DFG minus CFG: $0.0000$ $[-0.039,+0.043]$;
\item CFG+DFG minus no edges: $-0.0043$ $[-0.028,+0.020]$.
\end{itemize}
At seed 44, CFG versus no edges has 1 gain and 8 losses ($p=0.0391$), while adding DFG to CFG has 7 gains and 0 losses ($p=0.0156$). These opposite single-seed directions do not establish stable effects.

\subsection{Component Interventions}

\begin{table}[!t]
\centering
\caption{Seed-42 component interventions. The table is exploratory and not used in the main conclusions.}
\label{tab:architecture_components}
\begin{tabular}{lrr}
\toprule
Intervention & $\Delta\passat{10}$ & $p$ \\
\midrule
No-GINE vs. no-edge GINE & +.0065 & 1.000 \\
No-GINE vs. CFG+DFG GINE & +.0260 & 0.219 \\
CLAP-ASM vs. GraphCodeBERT encoder & -.0065 & 1.000 \\
Four block vectors vs. one CLS & +.0000 & 1.000 \\
Eight block vectors vs. one CLS & -.0130 & 0.688 \\
Hierarchical regions vs. block prefix & -.0195 & 0.375 \\
Region maximum 16 vs. no regions & +.0130 & 0.688 \\
No global attention & -.0130 & 0.625 \\
No block positions & -.0195 & 0.375 \\
Two- vs. four-layer GINE & +.0065 & 1.000 \\
Cross-task permuted graph & -.0130 & 0.500 \\
Shuffled block order & -.0065 & 1.000 \\
Zero prefix gate & -.0455 & 0.016 \\
Three additional sealed contrasts & -.0065 / .0000 / -.0195 & 1.000 / 1.000 / .508 \\
\bottomrule
\end{tabular}
\end{table}

With 16 contrasts, the smallest-rank Holm threshold is .003125; the prefix-gate result does not survive. If 16 null tests were independent, the probability of at least one nominal $p<.05$ would be about .56; dependence prevents treating that illustration as the actual family error rate. Zeroing the prefix gate has 0 gains and 7 losses, but this single contrast does not establish channel use after correction. Freezing the encoder reduces compile@1 from .78 to .48. Cross-task graph substitution and block-order shuffling have smaller observed effects. These lower-provenance diagnostics are not replications of the reviewer-requested instruction-content interventions in Section~\ref{sec:assembly_ablation}. On a high-complexity stratum (25--458 blocks), the no-edge/CFG/CFG+DFG arms score .180/.167/.160, versus .340/.320/.327 on the 3--14 block stratum; this is descriptive.

\subsection{Provenance Status}

Three of five planned core seeds were completed, and component isolations remain single-seed diagnostics. The provenance audit header reports \texttt{FAIL} across 38 runs because two CLAP runs record encoder revisions that differ from the expected revision. Separately, 10/38 aggregate summaries disagree with candidate-level replay, which is treated as canonical. A stability audit re-executed 3,450 archived passing candidates; 3,449 remained passing and one was invalidated. The appendix records these diagnostics and corrections, while architectural conclusions remain excluded from the main findings.

\section{Frontier-Model Lineage and Reference Point}
\label{app:frontier}

An initial frontier measurement of 5/60 was invalidated because assembly was truncated at 12,000 characters, 33/60 prompts lost the function tail, anonymous closures were omitted by the symbol regex, the cache was not content-addressed, and silent failures left effective $k$ unknown.

The replacement harness pins a paired 175-task manifest and enforces the following invariants: prompts exclude gold and tests; no prompt truncation; only verified seals are measured; prompt artifact hashes match; acceptance tests are identical; F2 grammar is identical; ordered tasks are identical; and the prompt-budget contract is identical. The evaluator, Dart binary, tokenizer, and prompt artifacts are SHA-256 pinned, each run carries a completion attestation, and incomplete runs produce no score.

The complete DeepSeek-V4-Pro run covers 175/175 tasks and 1,750/1,750 slots. It solves 4 tasks at \passat{10} (2.29\%). Task-level \compileat{10} is 86/175 (49.1\%); candidate-level compilation is 300/1,750 (17.1\%), and candidate-level passing is 11/1,750 (0.63\%). Every slot reports \texttt{finish\_reason: stop}. A fine-tuned 8B student reaches \compileat{10} of 69.1\% on the same tasks. This is one provider on one date and one decoding configuration, not a performance ceiling. Other queried models reached preflight or quota only; absence of a score is not a score of zero.

\section{Preliminary ARM64 Flutter Observation}
\label{app:arm64}

The existing FL-343 run evaluates held-out functions compiled into obfuscated, signed Flutter release APKs from model-generated Dart sources. It reports \passat{10}=5.54\% (19 tasks) and \compileat{5}=93.07\%. Success falls monotonically with instruction length; none of 168 functions with at least 200 instructions is solved, while \compileat{1} declines from approximately 0.65 to 0.39 across length bins.

The infrastructure provides four validated signature views for all 343 rows, with zero residual names in the scrubbed views and assembly byte-identical to the unmodified arm. Signature text and binary evidence enter physically separate channels, allowing exact binary removal. Baseline compilation is 339/343, leaving little upward headroom. The reported run uses the original semantic signature, one seed, and model-generated sources. It is therefore a preliminary cross-ISA and length-scaling observation, with the infrastructure ready for metadata-scrubbed, repeated-seed evaluation rather than a completed generalization test on non-synthetic code.

\section{Claim Ledger and Artifact Map}
\label{app:claim_ledger}

\begin{longtable}{p{0.22\textwidth}p{0.15\textwidth}p{0.50\textwidth}}
\caption{Claim ledger for the integrated article.}\label{tab:claim_ledger}\\
\toprule
Evidence & Status & Interpretation \\
\midrule
\endfirsthead
\toprule
Evidence & Status & Interpretation \\
\midrule
\endhead
Six historical fine-tuning comparisons & Primary, configuration-specific & No tested variant has a significant functional gain after Holm correction; Qwen3-8B variants regress. One training outcome per historical variant and a legacy forced-zero task limit scope. \\
Original separate-panel metrics & Motivating & Opposite aggregate movements occur on different tasks/candidates; they do not estimate same-candidate associations. \\
Three-seed graph-prefix retraining and aligned metric audit & Replicated secondary, provenance-limited & Seeds 42--44 create distinct checkpoints. On their common candidate pools, \codebleu has modest and compilation weak association with execution; candidate replay is canonical. \\
HE benchmark certification & Primary audit & The historical 154-task corpus contains distinct reference and contract defects; clean subsets are reported. \\
Semantic-metadata/name-only intervention & Primary, single-seed & Large within-run dependence on a metadata bundle; the explicit opaque name-only/binary arm solves 0/154; no seed-level stability claim. \\
Semantic-signature-only/no-binary comparator & Secondary & No detectable advantage for tested binary-conditioned arms; adaptation also differs; no equivalence or channel isolation. \\
Graph-encoder instruction shuffle and cross-task substitution (\S\ref{sec:assembly_ablation}) & Secondary, single-seed; completed ablations & Semantic HE-154 view: six-test Holm $p\geq.750$, not equivalence. Opaque name-only view: no passes in any condition; the observed floor prevents measuring further loss. Mapping hashes and consumed graph inputs are verified. No positive instruction-use attribution; typed-opaque perturbations and repeated seeds remain untested. \\
MF reseeding & Replicated secondary & Protocol-specific variation and task-set turnover. \\
Typed-contract MF effect & Supportive secondary & Six positive observed blocks; sixth selected prospectively after five were seen. \\
Constants and structure channels & Exploratory & Directional/unmatched results; non-significance is not equivalence. \\
No-binary typed oracle & Exploratory oracle & Interface conformance can saturate without correctness; not deployable. \\
Cross-system table & Descriptive & Mixed systems; no causal inference. \\
Rejection sampling and expanded SFT & Exploratory & No robust improvement; one diversity-collapse mode. \\
RL and dense distillation & Feasibility & No efficacy measurement was produced. \\
Architecture sweep & Exploratory / provenance-failed & Preserved in Appendix~\ref{app:architecture}; excluded from conclusions. \\
Initial frontier run & Withdrawn & Invalid harness. \\
Replacement frontier run & Exploratory reference & One model/provider/date. \\
ARM64 Flutter run & Preliminary & Names retained and one seed; not a cross-ISA replication. \\
Historical scoring-test pools & Withdrawn & Prompt leakage; audit record only. \\
HumanEval task-160 forced zero & Legacy defect with sensitivity analysis & Excluding the shared zero preserves exact test decisions; worst-case missing counts preserve v5/v6 regressions. Absolute scores remain legacy; missing execution is not recovered. \\
\bottomrule
\end{longtable}

The artifact's README and file manifest identify retained, omitted, and missing material. The graph-ablation files and replay script supplement the immutable v5 evidence, preserving each analysis's original date and status. Historical candidates, the construction transcript, and a definitive v1--v6 run ledger remain unavailable; these gaps are distinct from the newly supplied replayable evidence.

\section*{Data Availability}

\ArtifactReleaseStatus

The published predecessor is the immutable v5 archive at \url{https://doi.org/10.5281/zenodo.22062791}. Its file is \texttt{paper\_artifact\_capsule\_v5\_20260823.zip}, SHA-256 \texttt{0a60e996\allowbreak{}58d76cf7\allowbreak{}a8c2d52d\allowbreak{}d7215ded\allowbreak{}a4edff4e\allowbreak{}fbda1d62\allowbreak{}51c70afe\allowbreak{}081eb5ca}. That record predates the new graph-ablation runs and is not the revision-specific deposit.

The companion v6 capsule retains the v5 evidence and adds six graph-ablation prediction pools and score files, both input views, evaluator-only HE tests, the matching source variants and commands, exact statistical replay, task-160 sensitivity analysis, and pinned public-source receipts. Its replay script verifies candidate, input, source, graph-mapping, and decoder-prompt hashes; it does not regenerate model outputs or rerun Dart. The no-GINE/regions16 checkpoint is publicly retrievable from the immutable Hugging Face repository revision \texttt{e44be594\ldots fb43838}; its LFS SHA-256 equals the \texttt{e8e87260\ldots c347fc} value in every run. The full retrieval URL and digest are in \texttt{checkpoint\_retrieval.json}; weights are not bundled.

Raw historical candidates, the original HumanEval-Dart construction transcript and rejection log, and a definitive v1--v6 run-to-configuration ledger remain unavailable. The earlier aligned raw predictions remain omitted under the v5 release policy, with their per-candidate scores and receipts retained. Newly supplied HE evaluator-only material is separated from inference inputs; it is not the internal MF holdback corpus. Licenses are file-scoped: authored code is Apache-2.0, documentation CC-BY-4.0, and benchmark-derived material remains subject to the bundled upstream notices. The capsule does not grant new rights over legacy material of unresolved provenance.

\section*{Generative AI Usage}

Perplexity Computer coordinated an additional manuscript and response review using Claude Fable 5.1, GPT-5.6 Sol, Gemini 3.1 Pro, and Grok 4.6, and assisted with claim calibration, archived-score consistency checks, and \LaTeX{} revision. This assistance did not generate new experimental candidates or reproduce training or Dart execution.

ChatGPT-5.5 with \texttt{xhigh} reasoning assisted the original construction of HumanEval-Dart by translating HumanEval implementations and tests into Dart and executing the translated tasks for validation; the exact historical conversation and execution transcript were not retained, and the independent certification in this revision reports the surviving defects. Claude Opus 5 and OpenAI Codex assisted with drafting and revision, analysis and audit scripts, and organization of the prompt-channel survey. Qwen, Kimi, OpenCode, ChatGPT, and GLM produced alternative draft text or adversarial manuscript reviews; their suggestions were selectively integrated after checks against the archived evidence. The authors formulated the research questions, designed and selected the experiments, determined the evidence taxonomy, interpreted the results, verified every citation and survey row against the primary material, regenerated reported statistics from archived artifacts, adjudicated every model suggestion, and take full responsibility for the article. The seed-47 extension was authored and sealed by the authors before that seed was generated, but after the first five blocks had been inspected, as disclosed in Section~\ref{sec:channel_integrated}.

\end{document}

%% file: artifact_release.tex
\providecommand{\ArtifactVersionDOI}{\url{https://doi.org/10.5281/zenodo.22650409}}
\providecommand{\ArtifactRecordURL}{\url{https://zenodo.org/records/22650409}}
\providecommand{\ArtifactSHA}{\texttt{44e5637b\allowbreak{}e1a72d50\allowbreak{}7675b4d5\allowbreak{}91c6434e\allowbreak{}bd6099fa\allowbreak{}9be6f2ac\allowbreak{}392d96ab\allowbreak{}6b034be2}}
\providecommand{\ArtifactReleaseStatus}{%
The revision-specific v6 evidence capsule was published on 7 September 2026
at \ArtifactRecordURL{} (version DOI: \ArtifactVersionDOI).
The deposited archive is 57,706,396 bytes; its SHA-256 is
\ArtifactSHA.
The release supports manifest-integrity checks and archived-outcome statistical
replay. These checks do not regenerate historical predictions, reproduce
training, rerun Dart tests, or attest the complete loaded checkpoint state.
Redistribution terms remain file-scoped as documented in the capsule.}